\documentclass[lettersize,journal]{IEEEtran}
\usepackage{amsmath,amsfonts}
\usepackage{array}
\usepackage[caption=false,font=normalsize,labelfont=sf,textfont=sf]{subfig}
\usepackage{textcomp}
\usepackage{stfloats}
\usepackage{url}
\usepackage{verbatim}
\usepackage{graphicx}
\usepackage{enumitem}
\usepackage{multirow}
\usepackage{booktabs}
\usepackage{makecell}
\usepackage{diagbox}
\usepackage{hhline}
\usepackage{pifont}
\usepackage{threeparttable}
\usepackage{algpseudocode}
\usepackage{xstring}
\usepackage{tabularray}
\usepackage[ruled,linesnumbered]{algorithm2e}
\usepackage{bbding}
\usepackage{tabularx}
\usepackage{soul,color,xcolor}

\newcommand{\plevel}[1]{%
	\IfEqCase{#1}{%
		{1}{$\triangle$}%
		{2}{$\triangle\triangle$}%
		{3}{$\triangle\triangle\triangle$}%
	}[]%
}%
\newcommand{\plevelstar}[1]{%
	\IfEqCase{#1}{%
		{1}{$\star$}%
		{2}{$\star\star$}%
		{3}{$\star\star\star$}%
	}[]%
}%

\makeatletter
\newcommand{\ssymbol}[1]{^{\@fnsymbol{#1}}}

\newif\ifrevision
 \revisionfalse

\ifrevision

\else

\fi

\begin{document}

\title{Leveraging Local and Global Knowledge Integration with Time-Frequency Calibrated Distillation for Speech Enhancement}


\author{Jiaming Cheng, Ruiyu Liang, Ye Ni, Chao Xu, Jing Li, Wei Zhou, Rui Liu, Bj\"orn W.\ Schuller, \IEEEmembership{Fellow, IEEE}, and Xiaoshuai Hao  
\thanks{This work was supported in part by the Natural Science Foundation of the Jiangsu Higher Education Institutions of China under Grant No.\,25KJB510010, the Open Foundation of Jiangsu Provincial Engineering Research Center for Intelligent Perception Technologies and Equipment under No.\,ITS202501 and the National Natural Science Foundation of China under Grant No.\,62472227 and No.\,61871213. \textit{(Corresponding author: Ruiyu Liang and Xiaoshuai Hao)}}
\thanks{Jiaming Cheng, Chao Xu and Jing Li are with the School of Computer Science, Nanjing Audit University, Nanjing 211815, China; Ruiyu Liang is with the School of Communication Engineering, Nanjing Institute of Technology, Nanjing 211167, China; Ye Ni is with School of Information Science and Engineering, Southeast University, Nanjing 210096, China; Wei Zhou is with the Cardiff University, CF10 3AT, United Kingdom. Rui Liu is with the Inner Mongolia University, Hohhot 010021, China; Bj\"orn Schuller is with the CHI -- the Chair of Health Informatics, TUM University Hospital, Germany and also with GLAM -- the Group on Language, Audio, \& Music, Imperial College London, UK. Xiaoshuai Hao is with the Xiaomi EV, 100085, China. (e-mail: 270777@nau.edu.cn; liangry@njit.edu.cn; niye@seu.edu.cn; 270174@nau.edu.cn; lijing@nau.edu.cn; zhouw26@cardiff.ac.uk; liurui\_imu@163.com; bjoern.schuller@imperial.ac.uk; haoxiaoshuai714@163.com)}
\thanks{Manuscript received xxxx, xxxx; revised xxxx, xxxx.}}

\markboth{Journal of \LaTeX\ Class Files,~Vol.~14, No.~8, August~2021}%
{Shell \MakeLowercase{\textit{et al.}}: A Sample Article Using IEEEtran.cls for IEEE Journals}


\maketitle

\begin{abstract}
	In recent years, complexity compression of neural network (NN)-based speech enhancement (SE) models has gradually attracted the attention of researchers, especially in scenarios with limited hardware resources or strict latency requirements. In this paper, we propose an intra-set and inter-set recursive fusion framework with time-frequency calibrated knowledge distillation (I\textsuperscript{2}SRF-TFCKD) for SE. Different from previous distillation strategies for SE, the proposed framework fully exploits the time-frequency differential information of speech while facilitating both local information focusing and global knowledge circulation. Firstly, we construct a collaborative distillation paradigm for intra-set and inter-set correlations. Within a correlated set, multi-layer teacher-student features are pairwise matched for calibrated distillation. Subsequently, we generate representative features from each correlated set through recursive fusion to form the fused feature set that enables inter-set knowledge interaction. Secondly, we propose a multi-layer interactive distillation based on dual-stream time-frequency cross-calibration, which calculates the teacher-student similarity calibration weights in the time and frequency domains respectively and performs cross-weighting, thus enabling refined allocation of distillation contributions across different layers according to speech characteristics. We conduct experiments on both single-channel and multi-channel SE datasets. Objective evaluations demonstrate that the proposed KD strategy consistently and effectively improves the performance of the low-complexity student model and outperforms other distillation schemes.
\end{abstract}


\maketitle

\section{Introduction}
\label{intro}

Speech enhancement (SE) aims to remove background noises and recover clean speech from noisy mixtures. As a perception front-end algorithm, its applications have already penetrated into diverse domains including such as intelligent terminals, industrial manufacturing, medical health, etc.~\cite{zheng2023sixty}. With the recent emergence of data-driven deep learning approaches, SE models have begun leveraging massive training data to characterize both speech signals, noise components, and their nonlinear relationships. Capitalizing on the representational learning capabilities of deep neural networks (DNNs), deep learning-based SE methods have achieved superior noise suppression performances and have attained state-of-the-art results in international SE challenges~\cite{reddy20_interspeech}, gradually becoming the mainstream SE solutions.

However, the superior performances of current NN-based SE methods often comes with substantial computational overhead. Top-ranked SE models in international challenges and public benchmarks typically have over 5M parameters and more than 10G floating-point operations (FLOPs) per second. Such excessive computational and memory requirements create deployment bottlenecks for edge devices, particularly in teleconferencing systems, Bluetooth headsets and hearing aids. Consequently, the exploration of SE models balancing performance with lightweight design has emerged as a research hotspot and trend~\cite{ZHU2025107805}.

Model compression is an effective solution, including pruning, quantization, and knowledge distillation (KD). However, pruning and quantization still heavily depend on the original model structure and hardware support. In contrast, KD transfers knowledge from a powerful teacher to a compact student, offering better generality and flexibility—thus becoming the technical path chosen in this paper. In recent years, KD techniques have achieved remarkable progress across a wide range of research fields~\cite{tang2024direct,gong2025beyond,11029670}. A common characteristic of these methods is that they fully leverage intermediate representations instead of relying only on output-level distillation. This observation has also been validated in the SE field. Recently, Wan et al.~\cite{wan23_interspeech} propose a hierarchical attention-based distillation framework for SE, exploring the utilization of intermediate representations. Similarly, Nathoo et al.~\cite{nathoo2024two} introduce a fine-grained similarity-preserving KD loss, which aims to match the student’s intra-activation Gram matrices to that of the teacher. However, existing methods have not yet explored frameworks that are specifically tailored to the inherent characteristics of the SE task.

In this paper, we design an intra-inter representation fusion with time-frequency calibration KD (I\textsuperscript{2}SRF-TFCKD) framework to further optimize the cross-layer supervision distillation paradigm. On the one hand, unlike previous approaches~\cite{nathoo2024two,cheng2024residual} that conduct cross-layer interactions within isolated correlation sets, we propose collaborative distillation across intra-set and inter-set correlations, effectively promoting knowledge flow throughout the model architecture. On the other hand, we introduce temporal and spectral domain cross-computation to derive multi-layer distillation calibration weights, leveraging time-frequency characteristics for targeted distillation optimization. The backbone network for distillation employs the dual-path dilated convolutional recurrent network (DPDCRN) model~\cite{cheng2023dual}, which won the SE track of the L3DAS23 challenge. The main contributions of this work are as follows:
\begin{itemize}
	\item We design a collaborative distillation paradigm for intra-set and inter-set correlations. Specifically, the SE model is partitioned into multiple correlated sets, and multi-layer teacher–student matching distillation is performed both within each set and across sets.
	\item We introduce a recursive fusion mechanism to generate representative features for each correlated set from both teacher and student, and then integrate them into a fused feature set. The recursive fusion preserves the critical information of each set while reducing redundancy in cross-set pairing.
	\item We propose multi-layer time-frequency cross-calibration, which independently computes teacher-student similarity calibration weights along temporal and spectral dimensions for cross-weighting, enabling refined allocation of distillation contributions across different layers.
	\item Extensive evaluations on public SE benchmarks demonstrate that our proposed I\textsuperscript{2}SRF-TFCKD framework for SE surpasses existing distillation methods and enabling the student model to achieve competitive enhancement results with low computational complexity.
\end{itemize} 

\section{Related work}
\label{sec_related}

\subsection{Time-frequency processing for SE}

Speech signals inherently exhibit temporal evolution characteristics (e.g., syllabic rhythm and prosody) and frequency properties (e.g., formant and harmonic structures). Time-domain processing captures the nonuniform variations of speech energy along the temporal axis, whereas frequency-domain representations are well suited to differentiating the distribution patterns of speech and noise. Consequently, the joint processing of time and frequency domains (T-F processing) has received increasing attention in recent SE research~\cite{zheng2023sixty, FAN2023508}.

On one hand, T-F processing has been deeply integrated into the architectural design of SE networks. Models based on convolutional–recurrent frameworks~\cite{hu20g_interspeech,zhao2022frcrn} utilize convolutional modules to extract local spectral patterns in the frequency domain, while recurrent units are used to capture temporal dependencies in the time domain. More recent designs~\cite{le21b_interspeech,NI2026103726} adopt parallel time–frequency dual-path branches, unfolding intermediate representations sequentially along the time and frequency axes and employing recurrent units to alternately process and integrate information across both domains. All these architectures explicitly emphasize the interaction between time and frequency domains, jointly preserving temporal continuity and spectral details.

On the other hand, T-F interaction has also been reflected in the design of loss functions for SE models. Xia et al.~\cite{xia2020weighted} introduce two mean-squared-error (MSE)-based spectral losses, where frame-level voice activity detection (VAD) in the time domain is used to separate the speech distortion and noise suppression terms. The modulation-domain loss~\cite{9414965}, formulated via the spectro-temporal modulation index (STMI), evaluates the integrity of processed speech within a perceptually motivated T-F modulation space. These T-F domain losses fully leverage speech-specific characteristics to formulate regression distance measures. Overall, T-F processing enables SE models to simultaneously capture temporal context and perceive fine spectral structures, thereby providing a promising direction for optimizing SE frameworks.

\subsection{Knowledge distillation for SE}

Knowledge distillation~\cite{hinton2015distilling} has been widely applied in the fields of computer vision and natural language processing. In the acoustic field, KD methods are initially adopted for classification tasks such as automatic speech recognition~\cite{yoon2023inter}. Subsequently, KD for regression tasks like text-to-speech (TTS) and SE has gradually emerged~\cite{9767637}. Early distillation studies for SE models primarily focus on minimizing output-level discrepancies between teachers and students, exemplified by elite sub-band distillation~\cite{hao20b_interspeech}. These works aim to align student models with teacher patterns at the learning objective level but overlook the utilization of intermediate representations containing richer information. 

To address this, a cross-layer similarity distillation framework~\cite{cheng2022cross} is proposed for SE models, which aligns teacher-student features via similarity loss. Similarly, multiple studies have adopted multi-layer distillation strategies to compress SE models, including bidirectional KD~\cite{CHEN2025103218} and two-step fine-grained similarity-preserving KD~\cite{nathoo2024two}. However, existing studies still suffer from two major limitations: redundant interference in multi-layer teacher–student interaction matching, and insufficient exploitation of the intrinsic T-F characteristics of speech. On the one hand, recent works attempt to reduce interference by weighting layer-wise contributions with attention mechanism~\cite{wan23_interspeech} or by adopting modularized knowledge review strategies~\cite{cheng2024residual}. Nevertheless, the global circulation and interaction of knowledge across modules are overlooked. On the other hand, although the SE baselines used in these studies typically process temporal and frequency information via separate streams, the distillation process does not account for differentiated analysis across the T-F domains. Recent studies such as dynamic frequency adaptive distillation~\cite{10890458} adjust the distillation targets according to speech frequency bands, while independent attention transfer mechanism~\cite{han2024distil} is designed for temporal and channel dimensions. These methods consider the influence of either the temporal or frequency domain in distillation, but joint T-F modeling is still missing. Given that speech and noise exhibit distinct distribution patterns in both the time and frequency domains, it is necessary to incorporate such characteristics into the knowledge transfer process. To address these limitations, this work mainly includes two innovative improvements: the time-frequency calibration strategy for distillation and the intra-inter set global knowledge transfer mechanism.

\begin{figure}[t]
	\centering
	\includegraphics[width=\linewidth]{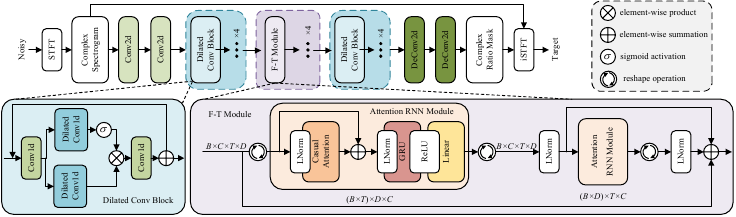}
	\caption{Backbone network architecture of the teacher model.}
	\label{fig_backbone}
\end{figure}

\section{Methodology}
\label{section_Methodology}

\subsection{Teacher and student network architecture}

\label{section_TS_network}

In this paper, we extend the multi-layer distillation within correlated sets~\cite{nathoo2024two,cheng2024residual} to intra-inter set global knowledge transfer, while implementing time-frequency cross-assignment weighting in the distillation loss. The backbone architecture chooses DPDCRN~\cite{cheng2023dual} that won the SE track of the L3DAS23 challenge, due to its facility for structural compression and scalability to both single-channel and multi-channel SE tasks. The backbone architecture of the teacher model is presented in Fig.\,\ref{fig_backbone}, where $B$ is the batch size, $C$ is the number of output channels, $T$ is the number of speech frames, $D$ is the feature dimension size. The input of the DPDCRN model is the complex spectrum obtained by applying the short-time Fourier transform (STFT) to the noisy speech. The output is the estimated complex ratio mask, which is multiplied with the noisy spectrum in the complex domain, followed by an inverse STFT (iSTFT) to reconstruct the enhanced speech. The main structure of DPDCRN consists of a convolutional encoder, a convolutional decoder, and an intermediate frequency-time (F-T) processing module. The encoder contains two convolutional layers and four dilated convolutional blocks, while the decoder contains two deconvolutional layers and four dilated convolutional blocks. Each frequency-time processing module contains dual branches for frequency and temporal axes, with each branch comprising a causal multi-head self-attention (MHSA) module and a feedforward network based on gated recurrent units (GRUs). Layer normalization (LNorm) and ReLU activation are incorporated between layers to regulate the distribution of feature flows. For each convolutional layer, the student has half the number of channels of the teacher. For the F-T processing modules, the number of hidden units in the student is half that of the teacher. Additionally, the teacher model has four cascaded F-T modules, while the student only uses one. Overall, the student contains merely 17\% of the teacher's parameters.

\subsection{Intra-inter representation fusion knowledge transfer architecture}

\begin{figure*}[t]
	\centering
	\includegraphics[width=0.84\linewidth]{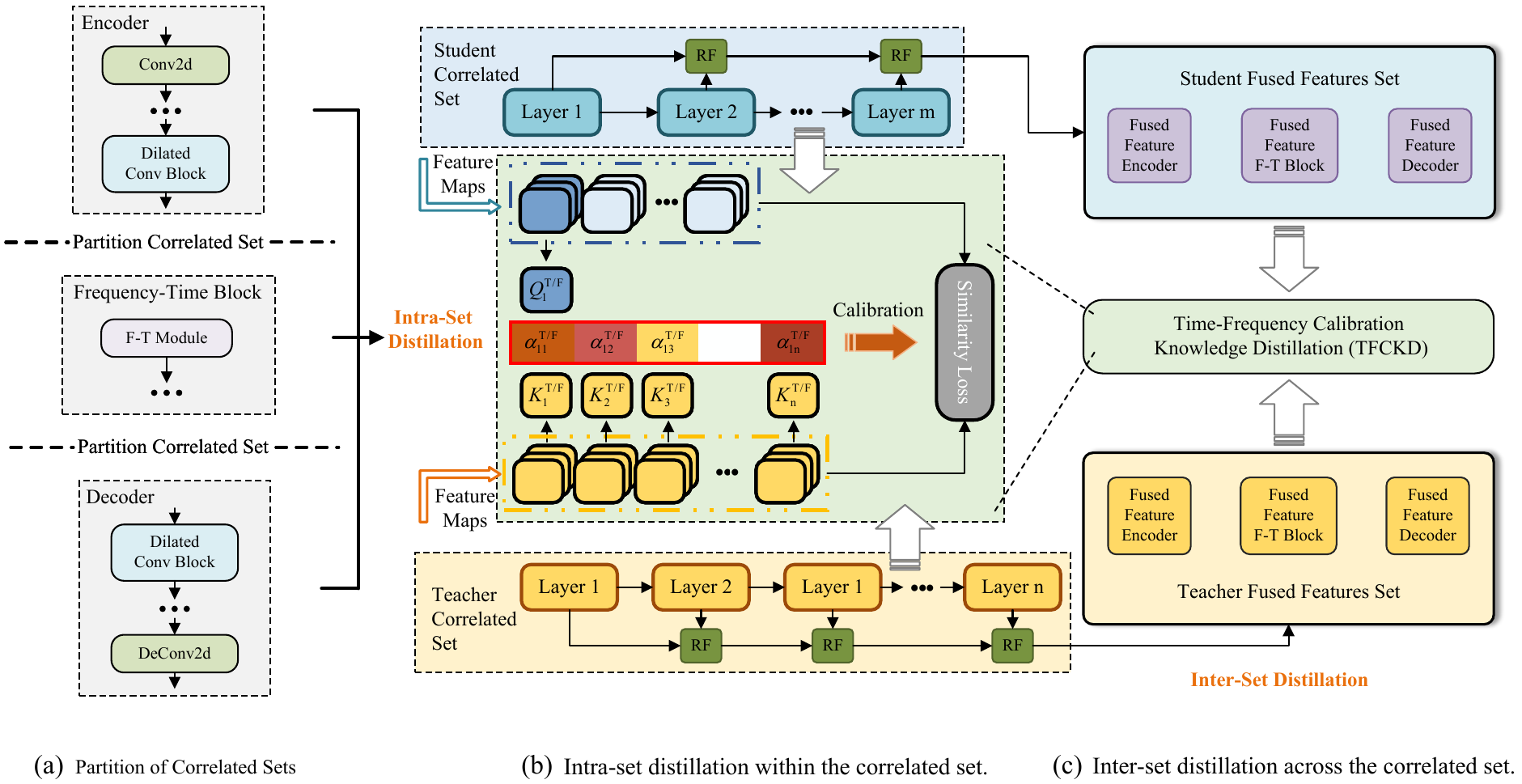}
	\caption{Overall architecture of the I\textsuperscript{2}SRF-TFCKD framework. We present the detailed process of intra-inter set distillation. Fig.\,\ref{fig_overall_arch}(a) denotes the partitioning of correlated sets. Fig.\,\ref{fig_overall_arch}(b) shows the time-frequency cross-calibration knowledge transfer and the recursive feature fusion across different layers within a single correlated set. Fig.\,\ref{fig_overall_arch}(c) demonstrates inter-set distillation achieved via the fused feature set among various correlated sets.}
	\label{fig_overall_arch}
\end{figure*}

According to the encoder-decoder framework of current mainstream SE models, we partition all layers of teacher-student networks into three correlated sets corresponding to the encoder, decoder, and intermediate F-T processing modules, respectively, to achieve hierarchical knowledge transfer. In previous work~\cite{cheng2024residual}, residual fusion is utilized to implement cross-layer knowledge transfer within each correlated set and confirmed its advantages over conventional layer-wise distillation. However, residual fusion distillation introduces redundant replication operations when the number of modules in teacher and student networks differs, and cross-set information interaction remains unexplored. In this paper, we propose a unified cross-layer distillation framework that jointly addresses intra-set and inter-set knowledge transfer. For intra-set operations, single-layer student features are aligned with all corresponding teacher representations through weighted matching within the correlated set. Regarding inter-set transfer, we first generate representative fusion features for each set via residual fusion, then perform multi-layer knowledge transfer on the fused feature set to facilitate cross-hierarchical circulation of intermediate representations. The overall architecture is shown in Fig.\,\ref{fig_overall_arch}.

\subsubsection{Intra-set multi-layer teacher–student feature distillation}

\begin{figure}[t]
	\centering
	\includegraphics[width=\linewidth]{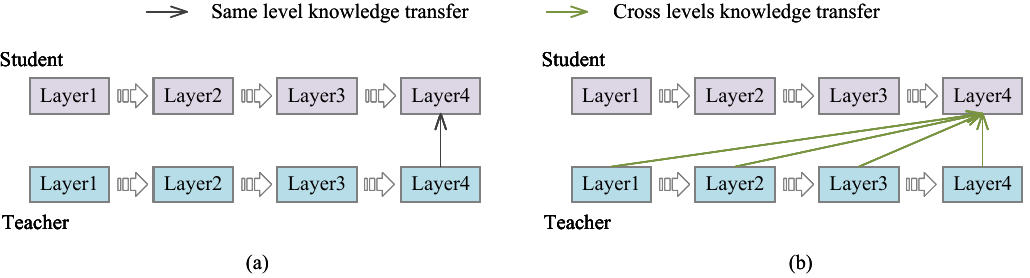}
	\caption{(a) using single-level teacher knowledge to guide one-level learning of the student. (b) using multiple layers of the teacher to supervise one layer in the student.}
	\label{fig_mkd}
\end{figure}

The candidate teacher-student correlation set is denoted as ${\cal P} = \left\{ {\left( {{l_s},{l_t}} \right)\left| {\forall {l_s} \in \left[ {1, \ldots ,{L_s}} \right]} \right.,{l_t} \in \left[ {1, \ldots ,{L_t}} \right]} \right\}$, where ${{l_s}}$ and ${{l_t}}$ represent the corresponding layers of the candidate student and teacher, respectively. As shown in Fig.\,\ref{fig_mkd}(a), single-layer feature distillation only transfers knowledge between teacher and student at the same hierarchical level:
\begin{equation}
	{{\cal L}_{singleKD}} = \sum\limits_{i \in {\cal P}} {{\cal D}\left( {T{r_s}\left( {{{\bf{F}}_{l_s^i}}} \right),T{r_t}\left( {{{\bf{F}}_{l_t^i}}} \right)} \right)} ,
\end{equation}
where ${{\bf{F}}_{{l_s}}}$ and ${{\bf{F}}_{{l_t}}}$ denote the hidden layer feature representations of the student and teacher respectively, $T{r_s}$ and $T{r_t}$ are used to transform teacher-student feature pairs into specific embedding representations for alignment. ${\cal D}$ is the computation of distillation distance. However, layer-wise distillation imposes strict constraints on structural and layer alignment between teacher and student models, requiring redundant layer information to be discarded in cases of mismatch. Therefore, we introduce multi-layer matching distillation for knowledge transfer within correlated sets. As illustrated in Fig.\,\ref{fig_mkd}(b), The single-layer student feature is aligned with hierarchical teacher features within the correlated set:
\begin{equation}
	\label{multiKD}
	{{\cal L}_{multiKD}} = \sum\limits_{\left( {{l_s},{l_t}} \right) \in {\cal P}} {{\cal D}\left( {T{r_s}\left( {{{\bf{F}}_{{l_s}}}} \right),T{r_t}\left( {{{\bf{F}}_{{l_t}}}} \right)} \right)} .
\end{equation}
Nevertheless, previous studies have shown that structural discrepancies across different functional modules may introduce redundant and interfering information during cross-layer distillation~\cite{cheng2022cross,cheng2024residual}. To address this issue, we partition the model into region-wise correlated sets and perform intra-set distillation to reduce interference. Specifically, the intra-set distillation explicitly accounts for the functional partitioning of SE models, where correlated sets are constructed in a one-to-one manner according to the encoder---T-F modeling---decoder processing pipeline. This design allows multi-layer matching to better align with the intrinsic characteristics of SE models, and will be further verified in Section~\ref{setion_sensitivity}. 

In addition, interactions between layers within the same correlated set may still introduce redundant information. To overcome this, we assign distillation weights across layers according to semantic information, which will be further discussed in Section~\ref{TFKD}.

\subsubsection{Inter-set distillation based on recursive representation fusion}

\begin{figure}[t]
	\centering
	\includegraphics[width=\linewidth]{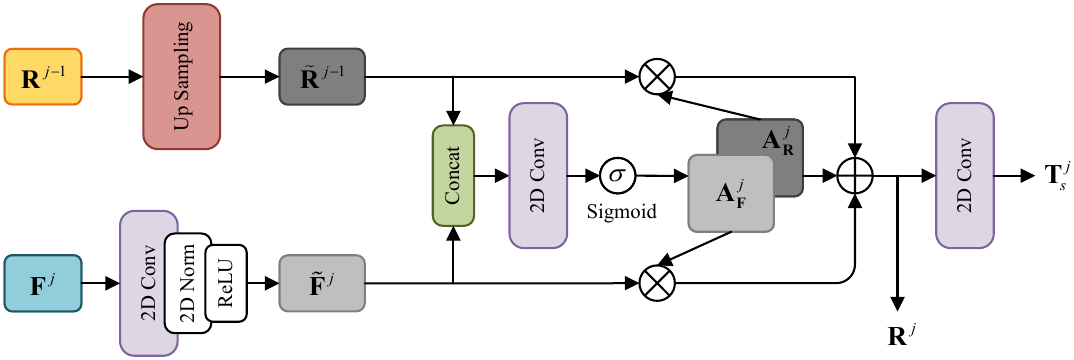}
	\caption{The process of generating fused features by utilizing the current layer features and the inherited recursive features.}
	\label{fig_fusion}
\end{figure}

Intra-set distillation, which performs teacher–student matching within each correlated set, encourages the student model to focus on fitting local functional representations, but it may overlook the cooperative relationships across different modules. As a result, the encoder---T-F modeling---decoder processing chain lacks a unified global constraint, which limits the overall distillation performance. Inspired by prototype distillation~\cite{10599808}, inter-set distillation is introduced to impose an additional constraint on global consistency across different functional modules. Specifically, we perform recursive fusion over each correlated set ${{\cal P}^k}$, which can be seen as a learnable aggregation operator ${\psi ^k}$:
\begin{equation}
	{{\bf{T}}^k} = {\psi ^k}\left( {{{\left\{ {{{\bf{F}}^l}} \right\}}_{l \in {{\cal P}^k}}}} \right),
\end{equation}
where ${{\bf{F}}^l}$ denotes the feature at the $l$-th layer within the $k$-th correlated set, and ${{\bf{T}}^k}$ represents the generated representative fused feature. Inter-set distillation is performed among these representative features, which can be regarded as imposing a summarized constraint across multiple correlated sets. Due to substantial scale discrepancies and training uncertainties among layer features from different correlated sets, their gradient directions may exhibit mutual interference. By leveraging recursive fusion, the features within each correlated set are aggregated into a function-level prototype representation. Aligning such prototypes enables the distillation to focus on the most stable and shared information of each correlated set, thereby facilitating effective teacher–student knowledge transfer along the global processing chain.

As illustrated in Fig.\,\ref{fig_overall_arch}(b), we first generate representative features for each correlated set through residual fusion to construct a teacher-student fused feature set. Then, the multi-layer distillation with time-frequency calibration is performed within this set. Fused features are generated via binary fusion of the current layer's features and recursive features are inherited from the previous layer, and are continuously propagated forward. The process of single-round residual fusion is shown in Fig.\,\ref{fig_fusion}.

The feature representation at the current $j$-th layer is denoted as ${{\bf{F}}^j}$, and the recursive features inherited from the previous layer are ${{\bf{R}}^{j - 1}}$. First, the representation dimensions are aligned via 2D convolution and up-sampling operations, respectively: 
\begin{equation}
	\begin{aligned}
		& {{{\bf{\tilde R}}}^{j - 1}} = {\rm{2D Conv}}\left( {{{\bf{R}}^{j - 1}}} \right),\\
		& {{{\bf{\tilde F}}}^j} = {\rm{Up Sampling}}\left( {{{\bf{F}}^j}} \right),
	\end{aligned}
\end{equation}
where the up sampling operation is used to transform the feature dimensions, while the 2D convolution is employed to adjust the channel dimensions. Subsequently, the transformed layer features and the recursive features are concatenated along the channel dimension, and an attention vector is generated via 1D convolution:
\begin{equation}
	{{\bf{A}}^j} = \sigma \left( {1{\rm{D Conv}}\left( {{\rm{Concat}}\left( {{{{\bf{\tilde F}}}^j},{{{\bf{\tilde R}}}^{j - 1}}} \right)} \right)} \right),
\end{equation}
where $\sigma$ represents the sigmoid activation function. The attention vector has two channels corresponding to the retention weight coefficients ${\bf{A}}_{\bf{F}}^j$ and ${\bf{A}}_{\bf{R}}^j$ for the layer features and recursive features, respectively. The recursive feature ${{\bf{R}}^j}$ of the current layer is generated by the weighted summation of the two branches:
\begin{equation}
	{{\bf{R}}^j} = {\bf{A}}_{\bf{R}}^j \cdot {{\bf{\tilde R}}^{j - 1}} + {\bf{A}}_{\bf{F}}^j \cdot {{\bf{\tilde F}}^j},
\end{equation}
where ${{\bf{R}}^j}$ will continue to participate in recursive computation as the fusion input for the next layer, while the fused output of the module is generated through 2D convolution:
\begin{equation}
	{{\bf{T}}^j} = {\rm{2D Conv}}\left( {{{\bf{R}}^j}} \right).
\end{equation}

For each correlated set, we select the last recursion output as the representative fused feature of the current set, which is then integrated to form a fused feature set. We believe that due to the symmetry of the encoder-decoder architecture in SE models, higher-level features from the middle layers exhibit stronger capability to learn effective information from lower-level features near the input/output ends. Therefore, in terms of residual fusion direction, the encoder and intermediate F-T processing modules follow a forward order starting from the first layer, while the decoder adopts a reverse order integrating from the final layer towards the middle. Corresponding representative features are generated for each correlated set in both the teacher and student models following the above process, and integrated into the teacher-student fused feature set for inter-set distillation.

\subsection{Time-frequency cross calibration for multi-layer hierarchical matching}
\label{TFKD}

\begin{figure}[t]
	\centering
	\includegraphics[width=0.92\linewidth]{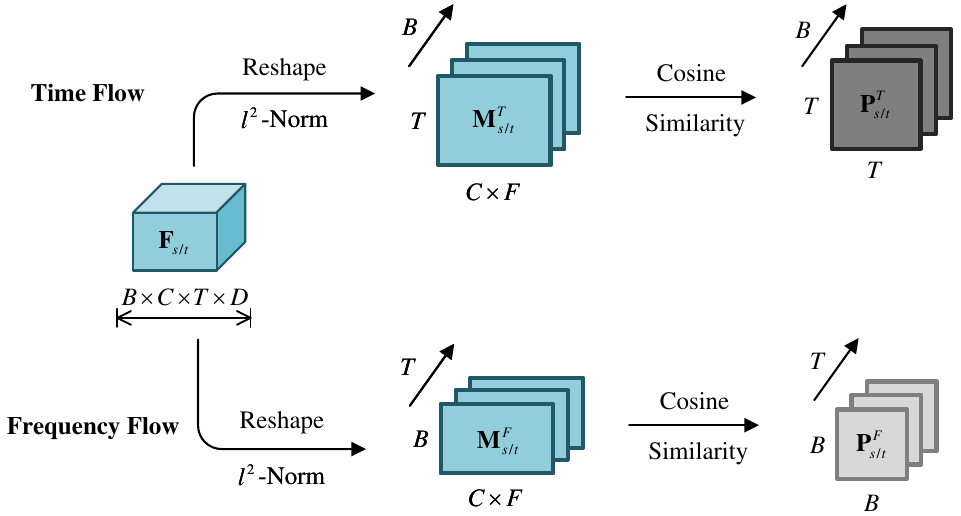}
	\caption{Similarity mapping of time and frequency flows. The self-similarity matrices for distillation are calculated in two flows: the time domain and the frequency domain. The distribution of features along the time axis will affect the similarity computation in the time flow, whereas the frequency domain processing operates frame-wise.}
	\label{fig_TF_Mapping}
\end{figure}

Different layer structures in DNNs exhibit varying capabilities to learn feature representations. As the depth of layers increases, the intermediate representation information of the model gradually becomes more abstract. Although multi-layer feature matching enables student models to perceive a broader range of teacher hidden knowledge, indiscriminate alignment may cause semantic redundancy that degrades learning efficiency and performance. To maximize absorption of valuable information, student layers should preferentially align with teacher layers exhibiting high semantic relevance. Inspired by the observation that the proximity of pairwise similarity matrices can be regarded as a good measurement of the inherent semantic similarity~\cite{wang2022semckd}, we use semantic-aware similarity for distillation calibration. Specifically, we introduce a calibration weight ${\alpha _{\left( {{l_s},{l_t}} \right)}}$ to dynamically adjust the contribution of candidate teacher-student layer pairs based on their semantic congruence, and constrain the corresponding weights through a softmax function: $\sum\limits_{{l_t} = 1}^{{L_t}} {{\alpha _{\left( {{l_s},{l_t}} \right)}}}  = 1,\forall {l_s} \in \left[ {1, \ldots ,{L_s}} \right]$. Eq.\,\ref{multiKD} after attention weighting is expressed as:
\begin{equation}
	{{\cal L}_{att - multiKD}} = \sum\limits_{\left( {{l_s},{l_t}} \right) \in {\cal P}} {{\alpha _{\left( {{l_s},{l_t}} \right)}}{\cal D}\left( {T{r_s}\left( {{{\bf{F}}_{{l_s}}}} \right),T{r_t}\left( {{{\bf{F}}_{{l_t}}}} \right)} \right)}.
\end{equation}

In fact, jointly considering frequency-band diversity and temporal dynamics is crucial for effective SE modeling. On the one hand, the distinct distributions of speech and noise across frequency bands provide key cues for noise suppression~\cite{chen2022fullsubnet+}. On the other hand, the temporal variation of speech and noise drive the models to capture dependencies ranging from local to long-term time scales~\cite{le21b_interspeech}. Given that SE models inherently perform separated computations in the time and frequency domains, differentiated analysis in the T–F domain should also be considered in distillation for SE. Therefore, we compute calibration weights for temporal and spectral flows independently and perform cross-weighting. Given a batch of feature map input ${{\bf{F}}_{s/t}} \in {{\mathop{\rm R}\nolimits} ^{B \times C \times T \times D}}$, where $B$ is the batch size, $C$ is the number of output channels, $T$ is the number of speech frames, $D$ is the feature dimension size, and $s/t$ denotes the student or teacher, we first calculate the similarity maps in the time-flow and frequency-flow respectively, as shown in Fig.\,\ref{fig_TF_Mapping}. For the time-domain flow, the feature map is split into multiple instances along the batch dimension, then flattened along the channel and feature dimensions. ${l^2}$ Normalization is applied to the merged feature dimension to obtain the temporal transformed features ${\bf{M}}_{s/t}^T \in {{\mathop{\rm R}\nolimits} ^{B \times T \times \left( {C \times D} \right)}}$. Finally, the time-flow mapping matrix ${\bf{P}}_{s/t}^T \in {{\mathop{\rm R}\nolimits} ^{B \times T \times T}}$ is computed using cosine similarity measurement:
\begin{equation}
	\label{similarity}
	{\bf{P}}_{s/t}^T = \frac{1}{2}\left( {\frac{{{{\left( {{\bf{M}}_{s/t}^T} \right)}^{\mathop{\rm T}\nolimits} } \cdot {\bf{M}}_{s/t}^T}}{{\left\| {{\bf{M}}_{s/t}^T} \right\|_2^2}} + 1} \right).
\end{equation}

For the frequency-domain flow, we partition the feature maps along the time dimension into multiple instances to maintain frame independence. Then, the frequency transformed features ${\bf{M}}_{s/t}^F \in {{\mathop{\rm R}\nolimits} ^{T \times B \times \left( {C \times D} \right)}}$ are obtained by flattening and normalization. Finally, the frequency-flow mapping matrix ${\bf{P}}_{s/t}^F \in {{\mathop{\rm R}\nolimits} ^{T \times B \times B}}$ is also calculated by cosine similarity.

Subsequently, we employ trainable weight matrices to project the teacher-student self-similarity mapping matrices from both the temporal and spectral domains into query and key subspaces, mitigating the effects of noise and sparsity:
\begin{equation}
	\begin{aligned}
		&{\bf{Q}}_{{l_s}}^{T/F} = EM{B_Q}\left( {{\bf{P}}_{{l_s}}^{T/F}} \right){\rm{,      }} \\ &
		{\bf{K}}_{{l_t}}^{T/F} = EM{B_K}\left( {{\bf{P}}_{{l_t}}^{T/F}} \right),
	\end{aligned}
\end{equation}
where $EM{B_Q}\left(  \cdot  \right)$ and $EM{B_K}\left(  \cdot  \right)$ consist of the fully connected layer, ReLU activation, and layer normalization. The time-domain and frequency-domain flows adopt independent trainable weight parameters. The similarity calibration coefficients for the time and frequency domains are computed through domain-specific query-key interactions: 
\begin{equation}
	\alpha _{\left( {{l_s},{l_t}} \right)}^{T/F} = {e^{{{\left( {{\bf{Q}}_{{l_s}}^{T/F}} \right)}^{\rm{T}}}{\bf{K}}_{{l_t}}^{T/F}}} \cdot {\left( {\sum\limits_{{l_t} \in {\cal P}} {{e^{{{\left( {{\bf{Q}}_{{l_s}}^{T/F}} \right)}^{\rm{T}}}{\bf{K}}_{{l_t}}^{T/F}}}} } \right)^{ - 1}}.
\end{equation}

Finally, cross-layer matching is performed for multiple teacher-student feature pairs within the current correlated set, where similarity metrics are computed through both time and frequency flows and distillation contributions are allocated via $T/F$ calibration coefficients. The multi-layer time-frequency calibration knowledge distillation (TFCKD) loss is calculated as follows:
\begin{equation}
	\begin{aligned}
		\mathcal{L}_{\mathrm{TFCKD}}
		&= \sum_{(l_s,l_t)\in\mathcal{P}}
		\Big[
		\alpha^{T}_{(l_s,l_t)}
		\mathcal{D}_{\mathrm{prop}}\!\left(
		\mathbf{P}_{l_s}^{T}, \mathbf{P}_{l_t}^{T}
		\right) \\
		&+
		\alpha^{F}_{(l_s,l_t)}
		\mathcal{D}_{\mathrm{prop}}\!\left(
		\mathbf{P}_{l_s}^{F}, \mathbf{P}_{l_t}^{F}
		\right)
		\Big].
	\end{aligned}
\end{equation}
where ${{\cal D}_{prop}}\left( {{\bf{P}},{\bf{Q}}} \right) = \left( {{\bf{P}} - {\bf{Q}}} \right)\log \left( {{{\bf{P}} \mathord{\left/
			{\vphantom {{\bf{P}} {\bf{Q}}}} \right.
			\kern-\nulldelimiterspace} {\bf{Q}}}} \right)$ represents the probabilistic distillation distance. ${\bf{P}}_{{l_s}/{l_t}}^T$ computes inter-frame feature correlations along the time axis, whereas ${\bf{P}}_{{l_s}/{l_t}}^F$ independently models frequency correlations within each frame. Consequently, T–F cross-calibration incorporates differentiated T-F analysis from both the perspectives of distance measurement and layer-wise contribution. This distillation loss serves as the knowledge transfer mechanism for both intra-set and inter-set teacher-student features.

\subsection{Training procedure}

\begin{algorithm}[t]
	\scriptsize
	\SetAlCapNameFnt{\footnotesize}
	\SetAlCapFnt{\footnotesize}
	\footnotesize
	\caption{Intra-inter distillation processing}
	\label{algo_I2RFKD}
	\DontPrintSemicolon
	\SetAlgoLined
	\SetKwFunction{Cal}{TF\_Calibration}
	\SetKwFunction{RF}{Fusion}
	
	\KwIn{Student and teacher correlated sets 
		$\{C_s^k\}_{k=1}^{q}$ and $\{C_t^k\}_{k=1}^{q}$.}
	\KwOut{Intra-set loss $\mathcal{L}_{\mathrm{Intra}}$ and inter-set loss $\mathcal{L}_{\mathrm{Inter}}$.}
	
	Initialize $\mathcal{L}_{\mathrm{Intra}} \leftarrow 0$, 
	$\mathcal{L}_{\mathrm{Inter}} \leftarrow 0$\;
	
	\For{$k \leftarrow 1$ \KwTo $q$}{
		Sample student features 
		$\mathbf{F}_s^k=\{\mathbf{F}_s^i\}_{i=1}^{m}$ from $C_s^k$\;
		Sample teacher features 
		$\mathbf{F}_t^k=\{\mathbf{F}_t^j\}_{j=1}^{n}$ from $C_t^k$\;
		
		\For{$i \leftarrow 1$ \KwTo $m$}{
			\For{$j \leftarrow 1$ \KwTo $n$}{
				$\mathcal{L}_{\mathrm{Intra}} \leftarrow 
				\mathcal{L}_{\mathrm{Intra}} + 
				\Cal(\mathbf{F}_s^i,\mathbf{F}_t^j)$\;
			}
		}
		
		$\mathbf{u}_s^k \leftarrow \RF(\mathbf{F}_s^1,\ldots,\mathbf{F}_s^m)$\;
		$\mathbf{u}_t^k \leftarrow \RF(\mathbf{F}_t^1,\ldots,\mathbf{F}_t^n)$\;
	}
	
	\For{$i \leftarrow 1$ \KwTo $q$}{
		\For{$j \leftarrow 1$ \KwTo $q$}{
			$\mathcal{L}_{\mathrm{Inter}} \leftarrow 
			\mathcal{L}_{\mathrm{Inter}} + 
			\Cal(\mathbf{u}_s^i,\mathbf{u}_t^j)$\;
		}
	}
\end{algorithm}

In this section, we elaborate on the entire training procedure for intra-inter distillation. The overall workflow is outlined in Algorithm~\ref{algo_I2RFKD}. First, we pretrain the teacher network using the multi-resolution short-time Fourier transform (MRSTFT) loss as the backbone loss. The teacher is then frozen to serve as a guide for the student, and KD proceeds concurrently with the student’s training. Intra-set distillation is applied to the encoder, decoder, and F-T processing modules. Within each correlated set, besides calculating the intra-set time-frequency calibration distillation loss ${{\cal L}_{Intra}}$, we generate representative teacher-student fused features for the current set through layer-wise recursive computation. Inter-set distillation loss ${{\cal L}_{Inter}}$ is derived via cross-layer interactions within the fused feature set. The total loss for training the student model is formulated as:
\begin{equation}
	\begin{aligned}
		\mathcal{L}_{stu}
		&= \mathcal{L}_{MRSTFT}
		+ \sum_{k=1}^{q}\sum_{i=1}^{m}\sum_{j=1}^{n}
		\mathcal{L}_{Intra}\!\left(
		\mathbf{F}_{s}^{k,i}, \mathbf{F}_{t}^{k,j}
		\right) \\
		&+ \sum_{i=1}^{q}\sum_{j=1}^{q}
		\mathcal{L}_{Inter}\!\left(
		\mathbf{u}_{s}^{i}, \mathbf{u}_{t}^{j}
		\right).
	\end{aligned}
\end{equation}
where $q$ is the total number of correlated sets (set to 3 in this paper), $m$ is the number of layers contained in the student correlated set, and $n$ is the number of layers contained in the teacher correlated set.

\section{Experimental settings}
\label{section_experimental_setup}

\subsection{Dataset setup}

The proposed distillation framework is trained and evaluated on two widely used public SE datasets: the single-channel deep noise suppression (DNS) challenge dataset\footnote{The dataset is available at \url{https://github.com/microsoft/DNS-Challenge}.} and the multi-channel L3DAS23 challenge dataset\footnote{The dataset is available at \url{https://www.l3das.com/icassp2023}.}. First, we conduct ablation studies on the DNS dataset to verify the effectiveness of the proposed distillation components. Subsequently, we compare our method with state-of-the-art (SOTA) models on the DNS official test set. Finally, the effectiveness of the proposed framework is further verified on the multi-channel L3DAS23 dataset.

The DNS dataset contains approximately 500 hours of clean speech and 180 hours of noise clips. We randomly partition the corpus into 60000 and 100 utterances for training and validation, respectively. Noisy speech samples are generated by mixing clean speech with noise at random signal-to-noise ratios (SNRs) ranging from -5dB to 15dB, utilizing 100 hours of speech data in total. For ablation studies, we construct a multi-SNR test set, where 100 speech clips that do not overlap with the training and validation sets are mixed with random noises at three SNR levels (-5dB, 0dB, and 5dB) to form 300 noisy-clean pairs. Additionally, the official non-reverb test set is used for comparisons of objective speech evaluation metrics between different algorithms.

The proposed distillation method's performance on the multi-channel SE task is tested in the 3D SE track of the L3DAS23 challenge. The speech data contains over 4,000 virtual 3D audio clips, each lasting up to 12 seconds. The noise set includes 12 categories of transient noise and 4 categories of continuous noise. The data are released as two first-order ambisonics recordings (each with 4 channels). We divide the training data that contains a total of 80 hours of noisy-clean speech pairs into training and validation sets at a ratio of 75:1. Since the official blind test set is not fully open-sourced, we compare various algorithms on the 7-hour development test set provided by the challenge. 

\begin{table}[!t]
	\renewcommand{\arraystretch}{1}
	\centering
	\caption{Baseline model hyperparameter settings.}
	\label{Base_Hyperparameter}
	\resizebox{\linewidth}{!}{
		\begin{tabular}{ccccc}
			
			\cmidrule[\heavyrulewidth](r){1-5}
			\multirowcell{2.5}{Layer} & \multicolumn{2}{c}{Teacher} & \multicolumn{2}{c}{Student} \\
			\cmidrule(r){2-3} \cmidrule(r){4-5}
			& Parameters & Output Size & Parameters & Output Size \\
			\cmidrule[\heavyrulewidth](r){1-5}
			
			Conv2d 
			& (1, 3), (1, 2), 128 
			& (128, $T$, $F/2$) 
			& (1, 3), (1, 2), 64 
			& (64, $T$, $F/2$) \\
			
			Conv2d 
			& (1, 3), (1, 2), 128 
			& (128, $T$, $F/4$) 
			& (1, 3), (1, 2), 64 
			& (64, $T$, $F/4$) \\
			
			\multirowcell{2}{Dilated Block\\(4$\times$Conv2d)} 
			& \multirowcell{2}{(2, 3), (1, 1), 128\\$p$: [1, 2, 4, 8]} 
			& \multirowcell{2}{(128, $T$, $F/4$)} 
			& \multirowcell{2}{$(2, 3)$, $(1, 1)$, 64\\$p$: [1, 2, 4, 8]} 
			& \multirowcell{2}{(64, $T$, $F/4$)} \\
			\\
			
			Frequency-Time Block 
			& (128 GRU$\times$2)$\times$4 
			& (128, $T$, $F/4$) 
			& 64 GRU$\times$2 
			& (64, $T$, $F/4$) \\
			
			\multirowcell{2}{Dilated Block\\(4$\times$Conv2d)} 
			& \multirowcell{2}{(2, 3), (1, 1), 128\\$p$: [1, 2, 4, 8]} 
			& \multirowcell{2}{(128, $T$, $F/4$)} 
			& \multirowcell{2}{$(2, 3)$, $(1, 1)$, 64\\$p$: [1, 2, 4, 8]} 
			& \multirowcell{2}{(64, $T$, $F/4$)} \\
			\\
			
			DeConv2d 
			& (1, 3), (1, 2), 128 
			& (128, $T$, $F/2$) 
			& (1, 3), (1, 2), 64 
			& (64, $T$, $F/2$) \\
			
			DeConv2d 
			& (1, 3), (1, 2), 2 
			& (2, $T$, $F$) 
			& (1, 3), (1, 2), 2 
			& (2, $T$, $F$) \\
			
			\cmidrule[\heavyrulewidth](r){1-5}
			
		\end{tabular}
	}
\end{table}

\begin{table}[!t]
	\renewcommand{\arraystretch}{1}
	\centering
	\caption{Distillation module hyperparameter settings.}
	\label{Distill_Hyperparameter}
	\resizebox{\linewidth}{!}{
		\begin{tabular}{ccccc}
			
			\cmidrule[\heavyrulewidth](r){1-5}
			\multicolumn{5}{c}{Similarity Embedding Mapping} \\
			\cmidrule(r){1-5}
			\multirowcell{2.5}{Layer} & \multicolumn{2}{c}{Time Flow} & \multicolumn{2}{c}{Frequency Flow} \\
			\cmidrule(r){2-3} \cmidrule(r){4-5}
			& Parameters & Output Size & Parameters & Output Size \\
			\cmidrule(r){1-5}
			
			Linear 
			& $T * factor$ units 
			& ($B$, $T$, $T * factor$) 
			& $B * factor$ units 
			& ($T$, $B$, $B * factor$) \\
			
			ReLU 
			& - 
			& ($B$, $T$, $T * factor$) 
			& - 
			& ($T$, $B$, $B * factor$) \\
			
			Linear 
			& $T$ units 
			& ($B$, $T$, $T$) 
			& $B$ units 
			& ($T$, $B$, $B$) \\
			
			${l^2}$Norm 
			& - 
			& ($B$, $T$, $T$) 
			& - 
			& ($T$, $B$, $B$) \\
			
			\cmidrule(r){1-5}
			\multicolumn{5}{c}{Residual Fusion} \\
			\cmidrule(r){1-5}
			\multirowcell{2.5}{Layer} & \multicolumn{2}{c}{Teacher} & \multicolumn{2}{c}{Student} \\
			\cmidrule(r){2-3} \cmidrule(r){4-5}
			& Parameters & Output Size & Parameters & Output Size \\
			\cmidrule(r){1-5}
			
			Conv2d 
			& (3, 3), (1, 1), 128 
			& (128, $T$, $F_t^j$) 
			& (3, 3), (1, 1), 64 
			& (64, $T$, $F_s^j$) \\
			
			Up Sampling 
			& - 
			& (128, $T$, $F_t^j$) 
			& - 
			& (64, $T$, $F_s^j$) \\
			
			Concatenate 
			& - 
			& (256, $T$, $F_t^j$) 
			& - 
			& (128, $T$, $F_s^j$) \\
			
			Conv1d 
			& 1, 1, 2 
			& (2, $T$, $F_t^j$) 
			& 1, 1, 2 
			& (2, $T$, $F_s^j$) \\
			
			Sigmoid 
			& - 
			& 2$\times$(1, $T$, $F_t^j$) 
			& - 
			& 2$\times$(1, $T$, $F_s^j$) \\
			
			Conv2d 
			& (3, 3), (1, 1), $C_t^j$ 
			& ($C_t^j$, $T$, $F_t^j$) 
			& (3, 3), (1, 2), $C_s^j$ 
			& ($C_s^j$, $T$, $F_s^j$) \\
			
			\cmidrule[\heavyrulewidth](r){1-5}
			
		\end{tabular}
	}
\end{table}

\subsection{Implementation details}

In this paper, we use the DPDCRN model as our baseline, which takes complex spectrograms as input and outputs complex ratio masks. For the single-channel DNS dataset, the input channels are set to 2, while for the multi-channel L3DAS23 dataset, the input channels are set to 16. The hyperparameters of the teacher and student models are shown in Table\,\ref{Base_Hyperparameter}, where $T$ is the number of speech frames, and $F$ is the frequency dimension of speech features. For the encoder and decoder parts, the number of channels in each convolutional layer of the teacher is twice that of the student. In Table\,\ref{Base_Hyperparameter}, $p$ represents the dilation factor, and both the teacher and student models use four types of dilation receptive fields. For the intermediate part, the teacher model employs four F-T modules, while the student model uses only one F-T module, with the number of GRU hidden units also being half those of the teacher model's.

The hyperparameter settings for the trainable layers in the proposed distillation components are shown in Table\,\ref{Distill_Hyperparameter}. The trainable components contain the computation of query and key embedding matrices in time-frequency calibration, as well as the generation process of recursively fused features for each correlated set. In the time-frequency calibration, the similarity embedding mappings are computed separately from two flow directions, with each flow containing two fully-connected layers for dimension transformation. The dimension transformation coefficient \textit{factor} in Table\,\ref{Distill_Hyperparameter} is set to four in this paper. Residual fusion integrates the current layer features with the recursive features inherited from the previous layer via trainable parameters. In Table \ref{Distill_Hyperparameter}, $C_{t/s}^j$ represents the number of teacher-student channels in the current layer, $F_{t/s}^j$ denotes the feature dimension of teacher-student features in the current layer, and $F_{t/s}^{j - 1}$ denotes the feature dimension in the previous layer. The number of channels for recursive features of the teacher and student are set to 128 and 64, respectively. Notably, the trainable parameters of the distillation components are only involved in the training of the student model and do not impose additional burden during inference.

All speech signals are sampled at 16\,kHz and segmented into 2.5-second chunks. For all models, the window length is set to 32\,ms with a frame shift of 16\,ms, and the FFT size is 512 points. We employ the MRSTFT loss~\cite{defossez20_interspeech} as the backbone loss. The models are implemented using PyTorch. The training configuration includes a learning rate of 0.0006 with the Adam optimizer, a batch size of 8, and a total training duration of 20 epochs.

\subsection{Comparative methods}

We conduct multiple comparative experiments to comprehensively evaluate the proposed distillation framework for SE. For the comparative experiments on KD methods, we consider the following five frameworks: CLSKD~\cite{cheng2022cross} that leverages cross-layer similarity between the teacher and the student for knowledge transfer; UCLFWPKD~\cite{cheng2024residual}, a frame-weighting residual probabilistic KD framework; ABC-KD~\cite{wan23_interspeech} that enables adaptive learning of compressed multi-layer knowledge through layer-wise attention mechanisms; Two-Step KD~\cite{nathoo2024two} that adopts a two-stage distillation framework, first pre-training the student using only KD criteria to match teacher activation patterns, followed by further optimization through supervised training. MPMTNet-KD~\cite{11029670} explores cross-layer stage context correlation from multiple directions via a multiattentive perception and a multilayer transfer network.

Furthermore, we compare multiple state-of-the-art algorithms on the two benchmarks, respectively, to verify the competitiveness of the low-complexity student model after distillation. For the DNS benchmark, we select eight SE models for comparison. Specifically, NSNet~\cite{xia2020weighted} introduces two SE loss functions based on mean squared error. DTLN~\cite{westhausen20_interspeech} integrates analysis-synthesis methods through two cascaded networks. GTCRN~\cite{10448310} incorporates grouped strategies and leverages subband feature extraction modules to balance model complexity and enhancement performance. FastEnhancer~\cite{11464802} achieves streaming SE through a simple encoder-decoder architecture with RNNFormer blocks. DCCRN~\cite{hu20g_interspeech} reconstructs the convolutional and recurrent network structure using complex-valued computations. FullSubNet+~\cite{chen2022fullsubnet+} is an improved version of FullSubNet~\cite{hao2021fullsubnet}, strengthening frequency-band discriminability through multi-scale convolutions and channel attention. FRCRN~\cite{zhao2022frcrn} introduces a frequency-recurrent module applied to 3D convolutional feature maps along the frequency axis. CTS-Net~\cite{li2021two} introduces a two-stage SE network, estimating magnitude spectral information in the first stage and suppressing residual noise while correcting phase information in the second stage. 

For the L3DAS23 benchmark, we choose the following five comparative models. Neural Beamforming~\cite{ren2021neural} serves as the baseline of the challenge, combining a beamforming framework designed for distributed microphones with deep neural networks. EaBNet~\cite{li2022embedding} designs two modules: an embedding module that learns a 3D embedding tensor to represent time-frequency information and a beamforming module that derives beamforming weights to achieve filter-and-sum operations. LMFCA-Net~\cite{10889867} introduces time-axis decoupled fully-connected attention and frequency-axis decoupled fully-connected attention mechanisms to achieve lightweight multi-channel SE. CCA Speech~\cite{wang2023stream} is a U-Net-based streaming attention framework that ranked third in the L3DAS23 challenge. DeFT-AN~\cite{lee2023deft} incorporates three different types of blocks for aggregating information in the spatial, spectral, and temporal dimensions.

\subsection{Evaluation metrics}

In this paper, we use six objective speech quality evaluation metrics, including the perceptual evaluation of speech quality (PESQ)\footnote{The evaluation codes are available at \url{https://ecs.utdallas.edu/loizou/speech/software.htm}.\label{note1}}, short-time objective intelligibility (STOI)\footnote{The evaluation codes are available at \url{http://www.ceestaal.nl/code/}.\label{note2}}, scale-invariant signal-to-noise ratio (SI-SNR)\footref{note1}, and three mean opinion ratings on quality scales\footref{note1}: signal distortion (CSIG), the intrusiveness of background noise (CBAK), and overall quality (COVL). For all these metrics, higher scores indicate better performance. Additionally, on the L3DAS23 dataset, we use the official recommended word error rate (WER)\footnote{The evaluation codes are available at \url{https://github.com/l3das/L3DAS22}.\label{note3}} to evaluate SE effects for speech recognition purposes, where lower values indicate better performance.

\section{Experimental results and discussion}
\label{section_results_and_analysis}
\subsection{Ablation study}

\begin{figure}[t]
	\centering
	\includegraphics[width=\linewidth]{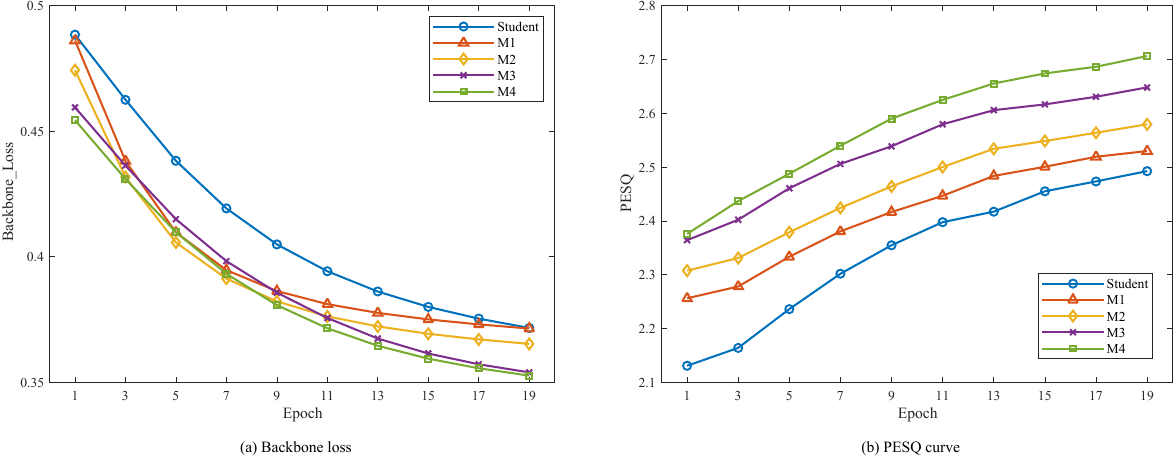}
	\caption{Training curve trends on the DNS validation set.}
	\label{fig_Training_Curve}
\end{figure}

We conduct ablation experiments for the distillation components on both the validation set and the multi-SNR test set to verify the effectiveness of the proposed modules step-by-step. We adopt layer-wise distillation with MSE loss as a distillation baseline. Building upon this, we introduce four configurations (M1–M4) to progressively validate the effectiveness of each component involved in the proposed I\textsuperscript{2}SRF-TFCKD framework. Specifically, M1 incorporates the layer-wise distillation with a similarity-based distillation loss, aiming to verify the effectiveness of similarity matrices in teacher-student representation distance measurement. Subsequently, M2 introduces intra-set cross-layer distillation over correlated sets, together with a basic attention weighting~\cite{wang2022semckd} for layer-wise distillation contributions. M3 further extends M2 by replacing the attention-based alignment with the proposed time–frequency cross-calibration described in Section~\ref{TFKD} of this paper. Finally, the M4 method (i.\,e., the proposed I\textsuperscript{2}SRF-TFCKD framework) performs time–frequency cross-calibrated multi-layer distillation both within and across correlated sets through recursive feature fusion, forming an intra–inter representation fusion distillation framework. The training trends of backbone loss and the speech quality evaluation metric PESQ on the validation set are illustrated in Fig.\,\ref{fig_Training_Curve}(a) and Fig.\,\ref{fig_Training_Curve}(b), respectively. Table\,\ref{ablation_metric} presents the comparison of objective speech metrics across different distillation strategies on the multi-SNR test set, and statistical significance analysis is conducted in the average metric comparison under three SNRs. We use the significance level $p$ to measure the statistical improvement of each distillation method compared to the student model.

\subsubsection{Sensitivity analysis for hyperparameters}
\label{setion_sensitivity}

\begin{table}[t]
	\caption{Averaged metrics across SNRs with different numbers of set partitions ($n$) for correlated sets.}
	\label{table_result_partition}
	\centering
		\resizebox{\linewidth}{!}{
		\begin{tabular}{lcccccc}
			\toprule
			
			Method & PESQ & STOI & SISNR & CSIG & CBAK & COVL \\
			\midrule
			Noisy & 1.405 & 0.821 &6.447& 2.423 &2.137& 1.865 \\ %
			\midrule
			\multicolumn{7}{l}{\bf Base model:} \\
			DPDCRN-T &2.628&0.946&14.596&4.069&3.415&3.376\\
			DPDCRN-S &2.264&0.910&13.821&3.701&3.173&2.986\\
			\midrule
			\multicolumn{7}{l}{\textbf{DPDCRN-S with intra-set multi-layer KD in \textit{n} correlated sets:}} \\
			\textit{n}=1  &2.325&0.914&13.863&3.731&3.232&3.055\\ 
			\multicolumn{7}{l}{\textit{n}=2}   \\ 
			\ \  Encoder \& F-T Block | Decoder &2.343&0.918&13.925&3.813&3.218&3.086 \\ 
			\ \ Encoder \& Decoder | F-T Block &2.343&0.915&13.917&3.777&3.232&3.067 \\ 
			\ \ Decoder \& F-T Block | Encoder &2.347&0.913&14.045&3.761&3.236&3.062 \\ 
			\textit{n}=3 (M2)  &\textbf{2.365}&\textbf{0.919}&\textbf{14.135}&\textbf{3.815}&\textbf{3.254}&\textbf{3.099}\\ 
			\textit{n}=\textit{number of all layers} (M1)  &2.332&0.914&13.997&3.765&3.218&3.056\\ 
			\bottomrule
		\end{tabular}
			}
	\vspace{-0.2cm}
\end{table}

\begin{table}[t]
	\caption{Averaged metrics across SNRs with different dimension transformation coefficients (\textit{factor}) of T-F calibration.}
	\label{table_result_factor}
	\centering
		\resizebox{\linewidth}{!}{
		\begin{tabular}{>{\raggedright\arraybackslash}p{4.8cm}cccccc}
			\toprule
			
			Method & PESQ & STOI & SISNR & CSIG & CBAK & COVL \\
			\midrule
			Noisy & 1.405 & 0.821 &6.447& 2.423 &2.137& 1.865 \\ %
			\midrule
			\multicolumn{7}{l}{\textbf{Base model:}} \\
			DPDCRN-T &2.628&0.946&14.596&4.069&3.415&3.376\\
			DPDCRN-S &2.264&0.910&13.821&3.701&3.173&2.986\\
			\midrule
			\multicolumn{7}{l}{\textbf{DPDCRN-S with intra-set multi-layer KD under T-F calibration (M3) with different \textit{factors}:}} \\
			\ \ \textit{factor}=1  &2.373&0.919&12.599&3.847&3.164&3.118\\ 
			
			\ \ \textit{factor}=2  &2.393&0.920&14.062&3.830&3.253&3.120\\ 
			\ \ \textit{factor}=4 (M3)  &\textbf{2.424}&\textbf{0.921}&\textbf{14.178}&\textbf{3.883}&\textbf{3.267}&\textbf{3.162}\\ 
			\ \ \textit{factor}=8  &2.404&0.919&13.696&3.845&3.235&3.135\\ 
			\ \ \textit{factor}=16 &2.414&0.919&13.791&3.866&3.252&3.150\\ 
			\bottomrule
		\end{tabular}
			}
	\vspace{-0.2cm}
\end{table}

\begin{table*}[h]
	\caption{Speech evaluation metrics for ablation comparisons of distillation components under different SNRs.}
	\label{ablation_metric}
	\centering
	\renewcommand{\arraystretch}{1}
	\begin{threeparttable}
		\resizebox{0.95\textwidth}{!}{%
			\begin{tabular}{lcccccccccccc}
				\toprule
				\multirow{3.5}{*}{Models} & 
				\multicolumn{4}{l}{PESQ} & 
				\multicolumn{4}{l}{STOI} &
				\multicolumn{4}{l}{SISNR} \\
				\cmidrule(lr){2-5} \cmidrule(lr){6-9} \cmidrule(l){10-13}
				& SNR &&&&SNR&&&&SNR&&& \\
				& -5\,dB & 0\,dB & 5\,dB & Avg & -5\,dB & 0\,dB & 5\,dB & Avg & -5\,dB & 0\,dB & 5\,dB & Avg \\ 
				\midrule
				Noisy &1.289&1.380&1.546&1.405&0.751&0.825&0.886&0.821&1.541&6.457&11.343&6.447\\ %
				\midrule
				
				\multicolumn{13}{l}{\bf Base model:} \\
				DPDCRN-T &2.249&2.644&2.992&2.628&0.935&0.940&0.964&0.946&11.744&14.717&17.327&14.596  \\
				DPDCRN-S&1.927&2.261&2.603&2.264&0.864&0.917
				&0.949&0.910&10.653&13.936&16.875&13.821
				\\
				\midrule
				\multicolumn{13}{l}{\bf DPDCRN-S with KD methods:} \\
				-Layer-wise KD \& MSE loss &1.939&2.299&2.673&2.304($\ast$)&0.866&0.918
				&0.950&0.911($\ast$)&10.595&13.921&16.913&13.810($\ast$)\\
				-Layer-wise KD \& Similarity loss (M1) &1.966&2.329&2.700&2.332($\ast$)&0.870&0.920
				&0.952&0.914($\ast$)&10.844&14.106&17.041&13.997($\ast$)\\
				-Intra-set KD \& Similarity loss (M2) &2.003&2.368&2.724&2.365($\ast$)&0.874&0.924
				&0.953&0.919($\ast\ast$)&11.020&14.237&17.149&14.135($\ast$)\\
				-Intra-set KD \& T-F calibration \& Similarity loss (M3) &2.057&2.426&2.788&2.424($\ast\ast$)&0.880&0.928
				&0.956&0.921($\ast\ast\ast$)&11.107&14.270&17.157&14.178($\ast$)\\
				-Intra-inter-set KD \& T-F calibration \& Similarity loss (M4 prop.) &\textbf{2.106}&\textbf{2.491}&\textbf{2.850}&\textbf{2.482}($\ast\ast\ast$)&\textbf{0.886}&\textbf{0.931}
				&\textbf{0.958}&\textbf{0.925}($\ast\ast\ast$)&\textbf{11.253}&\textbf{14.461}&\textbf{17.356}&\textbf{14.357}($\ast$)\\
				\midrule
				\midrule
				\multirow{3.5}{*}{Models} & 
				\multicolumn{4}{l}{CSIG} & 
				\multicolumn{4}{l}{CBAK} &
				\multicolumn{4}{l}{COVL} \\
				\cmidrule(lr){2-5} \cmidrule(lr){6-9} \cmidrule(l){10-13}
				& SNR &&&&SNR&&&&SNR&&& \\
				\cmidrule(l){2-13} 
				& -5\,dB & 0\,dB & 5\,dB & Avg & -5\,dB & 0\,dB & 5\,dB & Avg & -5\,dB & 0\,dB & 5\,dB & Avg \\ 
				\midrule
				Noisy &2.066 &2.407 &2.796 &2.423&1.828&2.108&2.475&2.137&1.612&1.844&2.140&1.865\\ %
				\midrule
				\multicolumn{13}{l}{\bf Base model:} \\
				DPDCRN-T &3.720&4.088&4.400&4.069&3.058&3.426&3.760&3.415&2.990&3.393&3.744&3.376\\
				DPDCRN-S &3.325&3.708&4.070&3.701&2.281&3.177
				&3.530&3.173&2.609&2.990&3.359&2.986\\
				\midrule
				\multicolumn{13}{l}{\bf DPDCRN-S with KD methods:} \\
				-Layer-wise KD \& MSE loss &3.353&3.752&4.125&3.743($\ast$)&2.821&3.203
				&3.573&3.199($\ast$)&2.629&3.031&3.425&3.029($\ast$)\\
				-Layer-wise KD \& Similarity loss (M1) &3.377&3.772&4.147&3.765($\ast$)&2.840&3.224
				&3.591&3.218($\ast$)&2.657&3.059&3.452&3.056($\ast$)\\
				-Intra-set KD \& Similarity loss (M2) &3.437&3.826&4.180&3.815($\ast\ast$)&2.884&3.262
				&3.617&3.254($\ast$)&2.708&3.107&3.483&3.099($\ast$)\\
				-Intra-set KD \& T-F calibration \& Similarity loss (M3) &3.500&3.904&4.244&3.883($\ast\ast\ast$)&2.894&3.274
				&3.633&3.267($\ast\ast$)&2.767&3.174&3.546&3.162($\ast\ast\ast$)\\
				-Intra-inter-set KD \& T-F calibration \& Similarity loss (M4 prop.) &\textbf{3.567}&\textbf{3.954}&\textbf{4.285}&\textbf{3.935}($\ast\ast\ast$)&\textbf{2.950}&\textbf{3.333}
				&\textbf{3.686}&\textbf{3.323}($\ast\ast\ast$)&\textbf{2.830}&\textbf{3.238}&\textbf{3.606}&\textbf{3.225}($\ast\ast\ast$)\\
				\bottomrule
			\end{tabular}%
		}
		\vspace{1mm}
		\begin{tablenotes}
			\footnotesize
			\fontsize{7pt}{8pt}\selectfont
			\parbox{0.75\linewidth}{\item[1] The significance analysis \cite{fisher1970statistical} of each distillation method and the baseline DPDCRN-S under average SNR is presented by the symbol $\ast$, and the significance level $p$ is: $\ast$: $p > 0.1$, $\ast\,\ast$: $0.05 < p < 0.1$, $\ast\ast\ast$ : $p < 0.05$.}
		\end{tablenotes}
	\end{threeparttable}
	\vspace{-0.2cm}
\end{table*}

The sensitivity analysis covers two aspects: the number of set partitions for correlated sets (\textit{n}) and the dimension transformation coefficient (\textit{factor}) in Table\,\ref{Distill_Hyperparameter}. The additional experiments are conducted on the multi-SNR test set, where we report the averaged metrics across multiple SNR levels.

On the one hand, regarding the partitioning of correlated sets, Table\,\ref{table_result_partition} reports the results of multi-layer KD in the M2 scheme on DPDCRN-S under four settings: \textit{n}=1, \textit{n}=2, \textit{n}=3, and \textit{n} equals the total number of layers. For fair comparison, the similarity loss is adopted as the distillation loss for all settings. Specifically, when \textit{n}=1, all layers of the teacher and student models are mutually matched. When \textit{n}=2, correlated sets are constructed by pairwise combinations of the three main modules---Encoder, F-T Block, and Decoder. The case \textit{n}=3 corresponds to the proposed partitioning strategy, where the correlated sets are defined according to the three main modules. When \textit{n} equals the total number of layers, multi-layer distillation degenerates into layer-wise distillation (M1). As shown in Table\,\ref{table_result_partition}, the case \textit{n}=1 yields inferior performances on most metrics compared with other partitioning settings. This indicates that layer representations from different modules in SE models may introduce information interference, and direct cross-module feature distillation can result in negative transfer. When \textit{n}=2, partitioned multi-layer distillation consistently outperforms layer-wise distillation across all metrics, demonstrating that module-wise intra-set distillation is more conducive to capturing key information. However, different pairwise combinations have only a marginal impact on the performance. The case \textit{n}=3, which partitions correlated-set aligned with the module-level architecture of the base model, achieves the best performance across all metrics. This indicates that the partitioning of correlated sets should have a direct correspondence with the model's functional regions.

On the other hand, for the dimension transformation coefficient (\textit{factor}) in Table\,\ref{Distill_Hyperparameter}, we report the performances of intra-set KD under T-F calibration with \textit{factor} varying in $\left[ {1,2,4,8,16} \right]$, as shown in Table\,\ref{table_result_factor}. As the \textit{factor} increases from 1 to 4, all speech metrics improve progressively, suggesting that moderately increasing the number of trainable parameters in the similarity embedding benefits the distillation process. However, when the \textit{factor} is further increased to 8 and 16, all performance metrics exhibit varying degrees of degradation, implying that excessive parameter expansion increases optimization difficulty and hinders teacher–student alignment. Therefore, the dimension transformation coefficient \textit{factor} is fixed to 4 in this work.

\subsubsection{The impact of cross-layer similarity distillation}

As shown by the results in Table\,\ref{ablation_metric}, incorporating the similarity loss into the distillation baseline leads to consistent performance improvements across all evaluation metrics for the student model. This indicates that implicitly reducing the distribution discrepancy between teacher and student through similarity matrices is more effective for SE distillation, and this design is therefore adopted in the subsequent distillation schemes. The effectiveness of correlated partitioning for intra-set distillation has been validated in Section~\ref{setion_sensitivity}. Here, we further examine the impact of cross-layer distillation on both the training loss curves and the evaluation metrics.

As shown in Fig.\,\ref{fig_Training_Curve}(a), M1 enables the student’s backbone loss to decrease rapidly in the early training stages, but this initial advantage does not persist until model convergence. In contrast, M2 achieves a lower converged backbone loss, indicating that cross-layer interactions provide additional knowledge to sustain the student’s learning. In the PESQ curves of Fig.\,\ref{fig_Training_Curve}(b), M2 consistently outperforms M1, which further validates this observation. In terms of the metric comparisons on the multi-SNR test set in Table\,\ref{ablation_metric}, M1 is inferior to other methods, showing no significant improvement ($p > 0.1$) over the student model in any of the six metrics. By contrast, M2 that extends distillation to multi-layer teacher–student interactions, outperforms M1 across all metrics and achieves secondarily significant improvements ($0.05 < p < 0.1$) in CSIG and STOI compared to the student model. These results demonstrate that the additional representational information provided by cross-layer interactions offers constructive guidance to the student in both speech quality and noise suppression.

\subsubsection{The impact of time–frequency cross-calibration}
Building upon the cross-layer KD framework of M2, M3 introduces the proposed time–frequency cross-calibration mechanism, which adjusts the distillation weights across layers tailored to the characteristics of speech signals. As shown in Fig.\,\ref{fig_Training_Curve}(a), the backbone loss of M3 exhibits a more uniform trend and achieves better convergence compared with M2, indicating that assigning multi-layer distillation weights based on time-frequency calibration leads to more stable knowledge transfer. Similarly, M3 also achieves a better performance ceiling on the PESQ curve of the validation set. Regarding the objective metrics on multi-SNR test set, M3 achieves statistically significant improvements over the student model on CSIG, COVL, and STOI ($p < 0.05$), and consistently outperforms both M1 and M2 across all three SNR conditions. These observations demonstrate that cross-calibration captures discriminative information across frequency bands and temporal segments at different layers, enabling more refined teacher-student layer matching.

\subsubsection{The impact of the intra-inter representation fusion framework}
Distinct from the previous three methods, M4 further introduces a global knowledge circulation mechanism across both intra-set and inter-set levels. This framework not only enables multi-layer teacher–student feature interactions within a single correlated set, but also facilitates cross-set deep fusion of teacher and student representations. As shown in the training curves of Fig.\,\ref{fig_Training_Curve}, M4 achieves the lowest convergence point of backbone loss among all schemes, while its PESQ curve consistently remains at the highest level, demonstrating stable and sustained performance gains. This indicates that intra-set and inter-set distillation complement each other, expanding the depth of knowledge perceived by the student model. Considering the objective metrics comparisons in Table\,\ref{ablation_metric}, M4 achieves statistically significant improvements ($p < 0.05$) over the student model on five metrics under average SNR, except for SISNR, and shows clear advantages over the other methods across all SNR levels. In summary, the proposed M4 method maximizes the transferability of the teacher by ensuring the student’s absorption of  intra-set fine-grained knowledge while simultaneously broadening its understanding of the teacher’s inter-set global information.

\begin{figure}[t]
	\centering
	\includegraphics[width=\linewidth]{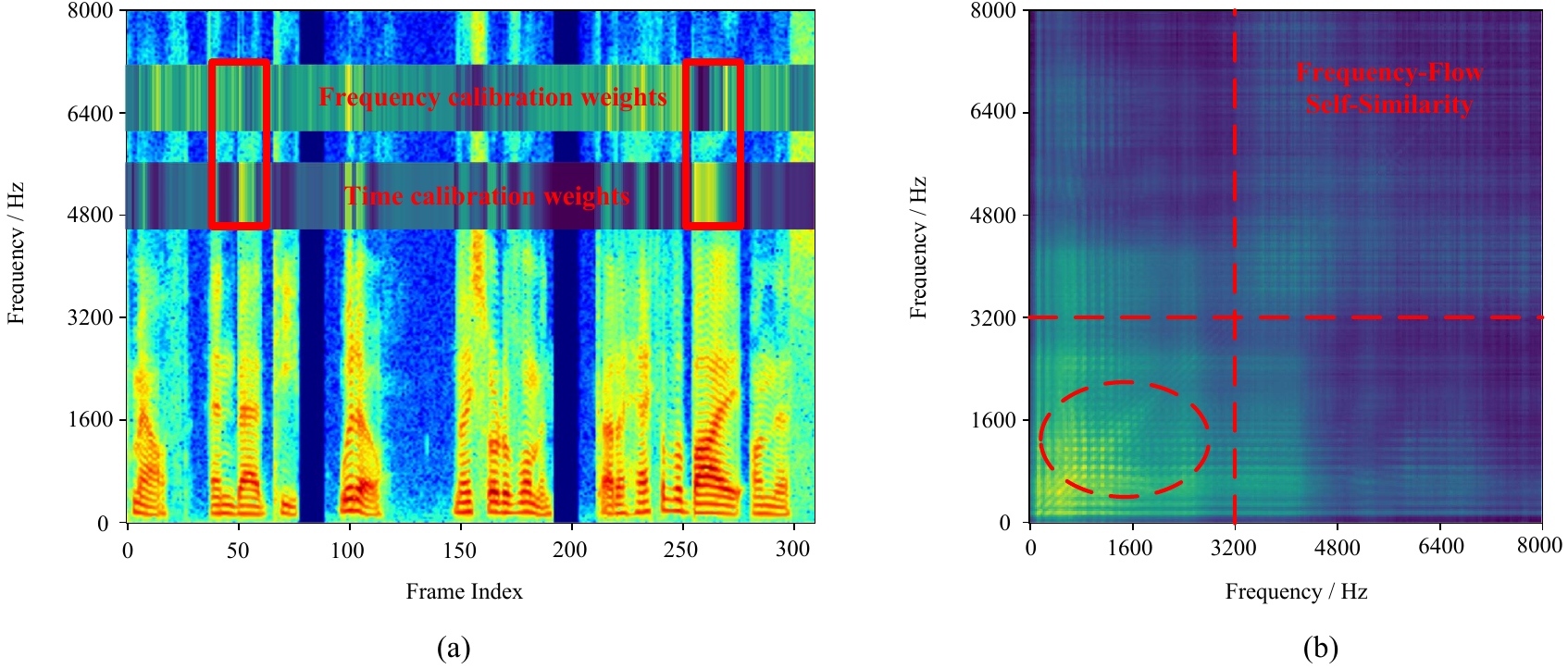}
	\caption{Heatmap distribution of T–F cross calibration. Panel (a) present heatmap visualizations of the calibration weights ${\alpha ^{T/F}}$ for both the temporal and spectral streams. Panel (b) visualize the self-similarity matrix of the frequency-transformed features ${{\bf{M}}^F}$.}
	\label{fig_heatmap}
\end{figure}

\begin{figure}[t]
	\centering
	\includegraphics[width=\linewidth]{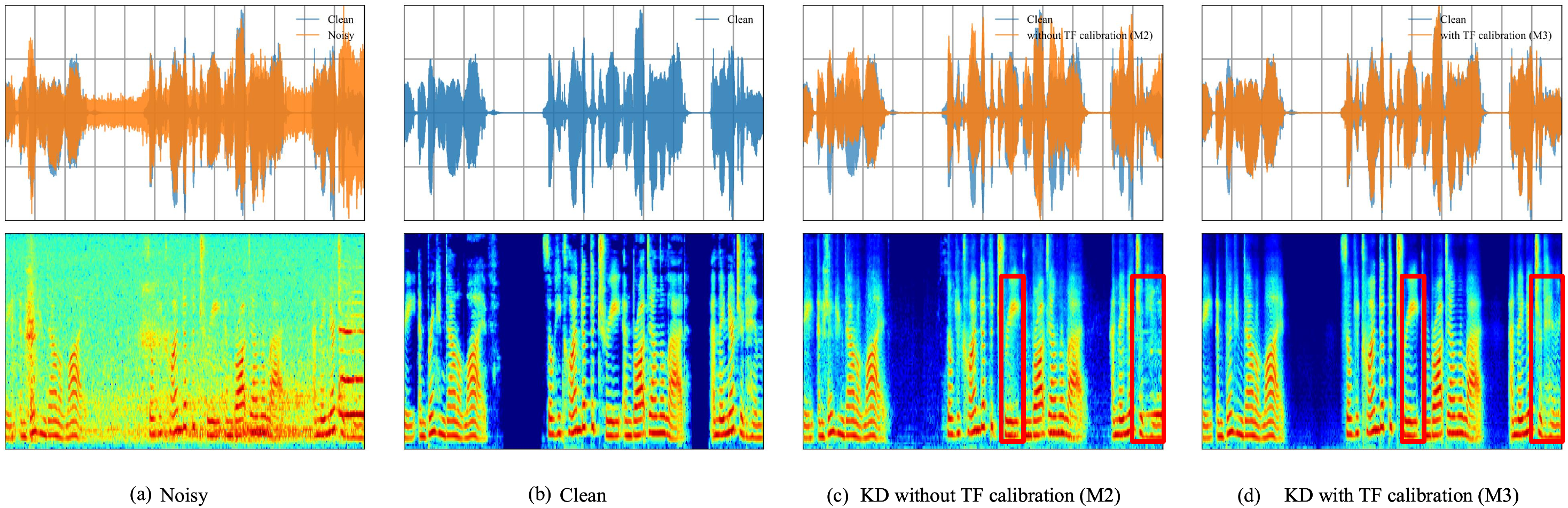}
	\caption{Comparison of distilled spectrograms with and without T-F calibration.}
	\label{fig_spectrogram}
\end{figure}

\subsubsection{Heatmap visualization analysis of time-frequency calibration weights}
We conduct heatmap and spectrogram analyses for the distillation with and without the proposed T–F calibration. In Fig.\,\ref{fig_heatmap}(a), we present heatmap visualizations of the calibration weights ${\alpha ^{T/F}}$ for both the temporal and spectral streams. We extract the trained time-frequency cross calibration weights for a speech segment and compute the frame-level average energy along the feature dimension. The temporal-stream weights are primarily concentrated in speech-active regions, whereas the spectral-stream weights show no clear focus. This observation arises because the temporal weight ${\alpha ^T}$ is computed along the time dimension, while the spectral weight ${\alpha ^F}$ is calculated independently for each frame. Therefore, the temporal calibration weights can capture contribution differences across different speech frames. Due to dimensional compression, the spectral weight ${\alpha ^F}$ cannot explicitly reflect the underlying frequency distribution. To better analyze the physical meaning of frequency calibration, we further visualize the self-similarity matrix of the frequency-transformed features ${{\bf{M}}^F}$ in Eq.\,\ref{similarity} which is used to compute ${\alpha ^F}$, as shown in Fig.\,\ref{fig_heatmap}(b). From the frequency distribution in the heatmap, the coupling is mainly concentrated in the low-frequency region with a boundary around 3.2 kHz, within which the energy is more compact. In contrast, the high-frequency region appears more dispersed, exhibiting weaker correlations and no clear focal structure. This observation is consistent with intrinsic speech characteristics, as the 1–3 kHz band contains the densest harmonic structures and plays a critical role in speech timbre and intelligibility. Such localized focusing enables the spectral stream calibration to concentrate on the frequency regions most relevant to speech.

In addition, we provide spectrogram analyses of multi-layer distillation with and without the proposed T–F calibration, as illustrated in Fig.\,\ref{fig_spectrogram}. Specifically, we present the time-domain waveforms and frequency-domain spectrograms of the noisy speech, clean speech, KD without T–F calibration (M2) and KD with T–F calibration (M3). From the comparison of the spectrograms, it can be observed that the model distilled with T–F calibration is able to suppress noise while preserving more spectral details, particularly the harmonic textures of speech. This observation is consistent with the improvements in speech quality metrics brought by the T–F calibration, indicating that the distillation process is enabled to better align with intrinsic speech characteristics.

\subsubsection{The trade-off between distillation performance and model compression ratio}
\begin{figure}[t]
	\setlength{\abovecaptionskip}{-0.2cm}
	\centering
	\includegraphics[width=\linewidth]{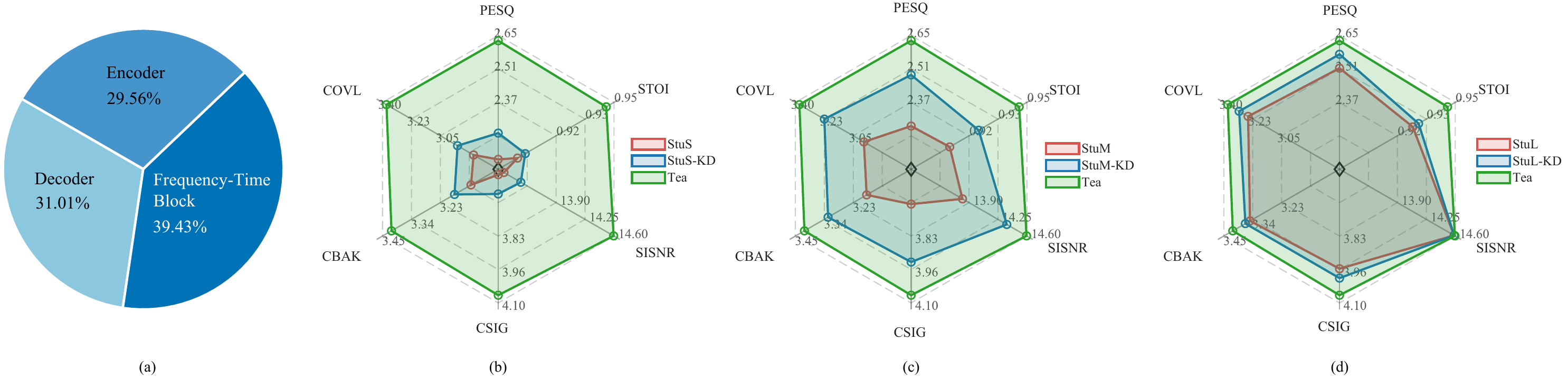}
	\vspace{0.5pt}
	\caption{Trade-off between model compression ratio and distillation performance. Panel (a) illustrates the FLOPs proportion of each module in the teacher model. Panels (b)–(d) respectively present the radar charts of student models with slim (S), medium (M), and large (L) complexity, distilled from the same teacher DPDCRN-T under the proposed I\textsuperscript{2}SRF-TFCKD framework.}
	\label{fig_performance_gap}
	\vspace{-0.2cm}
\end{figure}

\begin{table}[t]
	\caption{Distillation performances under different compression ratios.}
	\label{table_compression_ratio}

	\centering
	\begin{threeparttable}
		\resizebox{\linewidth}{!}{
			\begin{tabular}{lcccccccccc}
				\toprule
				
				Method & Compression ratio$^1$ & \makecell{Param.\\(M)} & \makecell{FLOPs$^2$\\(G/s)} & RTF$^3$ & PESQ & STOI & SISNR & CSIG & CBAK & COVL \\
				\midrule
				Noisy & - & - & - & - &1.405 & 0.821 &6.447& 2.423 &2.137& 1.865 \\ %
				\midrule
				DPDCRN-T & - & 3.5 & 13.71 & 0.36 &2.628&0.946&14.596&4.069&3.415&3.376\\
				DPDCRN-S (S) & 1/4, 1, 1/4 & 0.2 & 0.62 & 0.03 &2.122&0.900&13.269&3.582&3.106&2.849\\ 
				\ \ -I\textsuperscript{2}SRF-TFCKD & 1/4, 1, 1/4 & 0.2 & 0.62 & 0.03
				&2.234&0.904&13.473&3.660&3.170&2.946\\ 
				DPDCRN-S (M) & 1/2, 1, 1/2 & 0.6 & 2.44 & 0.09 &2.264&0.910&13.821&3.701&3.173&2.986\\ 
				\ \ -I\textsuperscript{2}SRF-TFCKD & 1/2, 1, 1/2 & 0.6 & 2.44 & 0.09
				&2.482&0.925&14.357&3.935&3.323&3.225\\ 
				DPDCRN-S (L) & 3/4, 2, 3/4 & 1.6 & 6.22 & 0.19 &2.511&0.928&14.586&3.962&3.348&3.253\\ 
				\ \ -I\textsuperscript{2}SRF-TFCKD & 3/4, 2, 3/4 & 1.6 & 6.22 & 0.19
				&2.570&0.931&14.513&4.000&3.366&3.307\\ 
				\bottomrule
			\end{tabular}
		}
		\begin{tablenotes}
			\fontsize{6pt}{7pt}\selectfont
			\parbox{0.55\linewidth}{\smallskip\item[1] Compression ratios for channels of convolutions, the number and hidden-layer dimension of F–T modules.}\\
			\parbox{0.55\linewidth}{\smallskip\item[2] Floating-point operations at the inference stage (same below).}\\
			\parbox{0.55\linewidth}{\smallskip\item[3] The real time factor (RTF) is measured with an Intel Core i5-12500 @ 2.50 GHz CPU by dividing the processing time by the input duration (same below).}
		\end{tablenotes}
	\end{threeparttable}

\vspace{-0.2cm}
\end{table}

The distillation metrics of the student model with different compression ratios under the same teacher model are analyzed in Fig.\,\ref{fig_performance_gap} and Table\,\ref{table_compression_ratio}. Fig.\,\ref{fig_performance_gap}(a) illustrates the FLOPs distribution across different parts of the teacher, where the F–T block accounts for the largest proportion. Accordingly, we reduce both the number of F-T modules and the dimensionality of the hidden features for the compression of the F–T block. Three compression levels are adopted, namely StuS, StuM, and StuL, with detailed configurations provided in Table\,\ref{table_compression_ratio}.

As shown in Fig.\,\ref{fig_performance_gap}(b)–(c) and Table\,\ref{table_compression_ratio}, StuS, which has the lowest complexity, exhibits a substantial performance gap from the teacher across all metrics. Although distillation brings some improvements, the gap remains significant. In contrast, StuL, with the highest complexity, achieves more than 97\% of the teacher’s metrics after distillation. However, due to its strong baseline performance, the relative gain from distillation is marginal. The student model with a moderate compression ratio, StuM, demonstrates more considerable improvements from distillation: its PESQ increases from 85\% of the teacher’s to 94\%, and the area of the six-dimensional radar plot is obviously expanded. Consequently, StuM achieves the best trade-off between performance and compression ratio and is therefore chosen as the baseline model for the distillation experiments in this paper.

\subsection{Comparative experiments on the DNS benchmark}

Fair comparisons of various algorithms are conducted on the DNS non-reverb test set. Table\,\ref{table_result_dns} presents the objective speech quality evaluation metrics of various SE methods. We provide details on the causality of each model and whether it uses future frame information (Casual and No Future Information, Casual \& NFI). Unlike the 40\,ms look-ahead constraint of the DNS challenge, we impose strict restrictions on both the causal architecture and the exclusion of future information to better adapt to real-time applications. Additionally, the model parameter (Param.) count and floating-point operations (FLOPs) per second are provided to evaluate deployment feasibility on edge devices. Among the SE models in Table\,\ref{table_result_dns}, NSNet, DTLN, GTCRN and FastEnhancer-L are low-complexity real-time SE algorithms. These models are fully causal and do not utilize future information. In contrast, DCCRN, FullSubNet+, FRCRN, and CTS-Net fail to satisfy both causality and NFI constraints. Although some architectures, such as DCCRN and CTS-Net, employ causal designs, they still leverage partial future frame information during training and inference. Our distillation baseline DPDCRN, composed of causal convolutions and GRU layers, is strictly causal and operates without any future frame information.

\subsubsection{Analysis of distillation frameworks}

We compare the effects of different distillation frameworks applied to the base model DPDCRN in Table\,\ref{table_result_dns}. Among these methods, M1, which performs layer-wise feature distillation, lags behind across most evaluation metrics. In contrast, CLSKD which incorporates cross-layer representation integration, achieves noticeable performance improvements. M2 further considers the partitioning of functional correlated sets on top of cross-layer distillation, leading to additional gains. Recent studies such as UCLFWPKD, ABC-KD, and Two-Step KD introduce targeted designs in cross-layer path connections and feature distance metrics, effectively enhancing the transmission of information and thus achieving considerable advantages in performances. Although MPMTNet-KD is not specifically designed for the SE task, it still achieves competitive results, which strongly demonstrates the importance of cross-layer context correlation in distillation for SE. In this paper, M3 introduces time-frequency cross-calibration based on the multi-layer distillation framework of M2, achieving notable improvements across all three metrics. This highlights that computing multi-layer calibration weights along both the temporal and spectral domains enables more effective knowledge allocation. Extending M3 to the intra-inter set distillation framework (M4, I\textsuperscript{2}SRF-TFCKD) achieves the best performance among all distillation methods, demonstrating that knowledge circulation within and between correlated sets further facilitate teacher-student knowledge transfer.

\begin{table}[t]
	\renewcommand{\arraystretch}{1.1}
	\caption{Comparison of objective speech evaluation metrics between distilled and undistilled models on the DNS test set.}
	\label{table_result_dns}
	\centering
	\begin{threeparttable}
		\resizebox{\linewidth}{!}{
			\begin{tabular}{lccccccc}
				\toprule
				Methods & Casual \& NFI & Param. (M) & FLOPs (G/s) & RTF & PESQ & STOI & SISNR (dB) \\
				\toprule
				Noisy & - & - & - &  - & 1.58 & 0.915 & 9.07\\ %
				\midrule
				NSNet~\cite{xia2020weighted}$\dagger$$^1$ & \Checkmark & 1.3 & 0.13 & 0.01 & 2.15 & 0.945 & 15.61 \\
				DTLN~\cite{westhausen20_interspeech}$\dagger$ & \Checkmark & 1.0 & 0.25 & 0.02 & - & 0.948 & 16.34 \\
				GTCRN~\cite{10448310}$\ddagger$ & \Checkmark & 0.05 & 0.04 & 0.07 & 2.64 & 0.955 & 16.84 \\
				FastEnhancer-L~\cite{11464802}$\ddagger$ & \Checkmark & 0.68 & 7.17 & 0.20 & 2.92 & 0.966 & 16.47 \\
				DCCRN~\cite{hu20g_interspeech}$\dagger$ & \XSolidBrush & 3.7 & 14.37 & - & 2.80 & 0.964 & 16.38 \\
				FullSubNet+~\cite{chen2022fullsubnet+}$\dagger$	& \XSolidBrush & 8.7 & 31.12 & - & 2.98 & 0.967 & 18.34 \\
				FRCRN~\cite{zhao2022frcrn}$\dagger$ & \XSolidBrush & 6.9 & 12.30 & - & 3.23 & 0.977 & 19.78 \\
				CTS-Net~\cite{li2021two}$\dagger$	& \XSolidBrush & 4.4 & 5.57 & - & 2.94 & 0.967 & 17.99 \\
				DPDCRN-T & \Checkmark & 3.5 & 13.71 & 0.36 & 3.16 & 0.973 & 17.47 \\
				DPDCRN-S & \Checkmark & 0.6 & 2.44 & 0.09 & 2.81 & 0.962 & 17.05 \\
				\hline
				\hline
				\addlinespace[4pt]
				KD methods &  &  &  &  &  & & \\
				\cline{1-1}
				\addlinespace[4pt]
				-CLSKD~\cite{cheng2022cross}$\dagger$ & \Checkmark & 0.6 & 2.44 & 0.09 & 2.84 & 0.963 & 17.08 \\
				-UCLFWPKD~\cite{cheng2024residual}$\dagger$ & \Checkmark & 0.6 & 2.44 & 0.09 & 2.94 & 0.966 & 17.16 \\
				-ABC-KD~\cite{wan23_interspeech}$\ddagger$ & \Checkmark & 0.6 & 2.44 & 0.09 & 2.90 & 0.964 & 17.14 \\
				-Two-Step KD~\cite{nathoo2024two}$\ddagger$ & \Checkmark & 0.6 & 2.44 & 0.09 & 2.93 & 0.965 & 17.05 \\
				-MPMTNet-KD~\cite{11029670}$\ddagger$ & \Checkmark & 0.6 & 2.44 & 0.09 & 2.92 & 0.964 & 17.08 \\
				-M1 & \Checkmark & 0.6 & 2.44 & 0.09 & 2.83 & 0.962 & 17.22 \\
				-M2 & \Checkmark & 0.6 & 2.44 & 0.09 & 2.89 & 0.965 & 17.36 \\
				-M3 & \Checkmark & 0.6 & 2.44 & 0.09 & 2.97 & 0.967 & 17.58 \\
				-M4(prop.) & \Checkmark & 0.6 & 2.44 & 0.09 & 3.03 & 0.968 & 17.74 \\
				
				\bottomrule
			\end{tabular}
		}
		
		\begin{tablenotes}
			\fontsize{6pt}{7pt}\selectfont
			\parbox{0.64\linewidth}{\smallskip\item[1] The model results labeled with symbol $\dagger$ are taken directly from the original paper, while the symbol $\ddagger$ denotes the results reproduced on the datasets of this paper according to the original paper's settings.}
		\end{tablenotes}
	\end{threeparttable}
\end{table}

\begin{figure}[t]
	\centering
	\includegraphics[width=\linewidth]{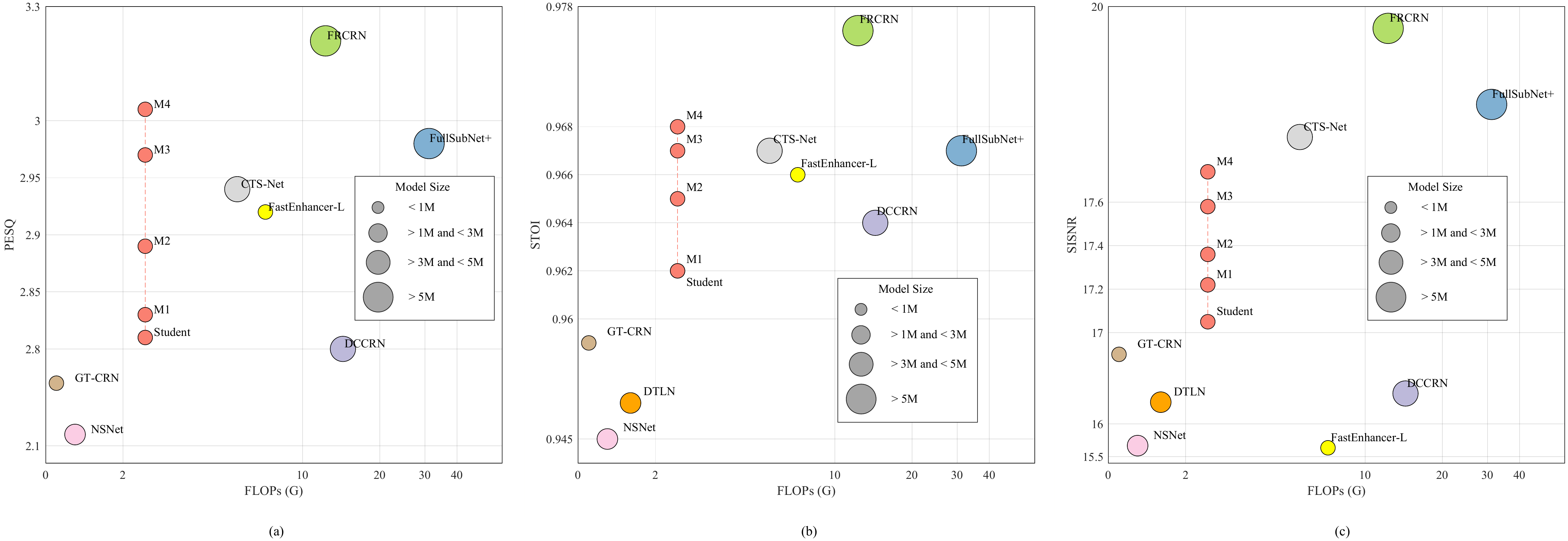}
	\caption{The scatter-bubble distribution of three metrics (PESQ, STOI, and SISNR) with model parameters and FLOPs for SE models on the DNS test set.}
	\label{fig_scatter_bubble}
\end{figure}

\subsubsection{Comparison with state-of-the-art SE methods}

Table\,\ref{table_result_dns} provides the metric performances of current SOTA SE algorithms on the DNS test set, while Fig.\,\ref{fig_scatter_bubble} shows the scatter-bubble distribution of three metrics (PESQ, STOI, and SISNR) with model parameters and FLOPs. In addition, for lightweight real-time SE models, we report the real-time factor (RTF), which is directly related to inference speed. It can be observed that current real-time SE algorithms such as NSNet, DTLN and GTCRN, despite having extremely low parameters and FLOPs, still exhibit a obvious performance gap compared to high-complexity SE models. The recently proposed real-time method FastEnhancer-L effectively narrows the gap with non-real-time methods, but it still involves relatively high FLOPs and shows unsatisfactory performance in terms of SISNR. Models like FRCRN and FullSubNet+ have metric advantages, but their high parameter counts (\textgreater 5M) and FLOPs (\textgreater 10G) restrict edge-side applicability. This paper applies intra-inter set distillation with time-frequency calibration (I\textsuperscript{2}SRF-TFCKD) to the real-time student model DPDCRN-S, achieving competitive performances while maintaining a low parameter count (0.6M), low computational complexity (2.4G FLOPs), and fast inference speed (0.09 RTF).

\subsection{Comparative experiments on the L3DAS23 benchmark}

\subsubsection{Boxplot analysis of distillation strategies}
\begin{figure}[t]
\centering
\includegraphics[width=\linewidth]{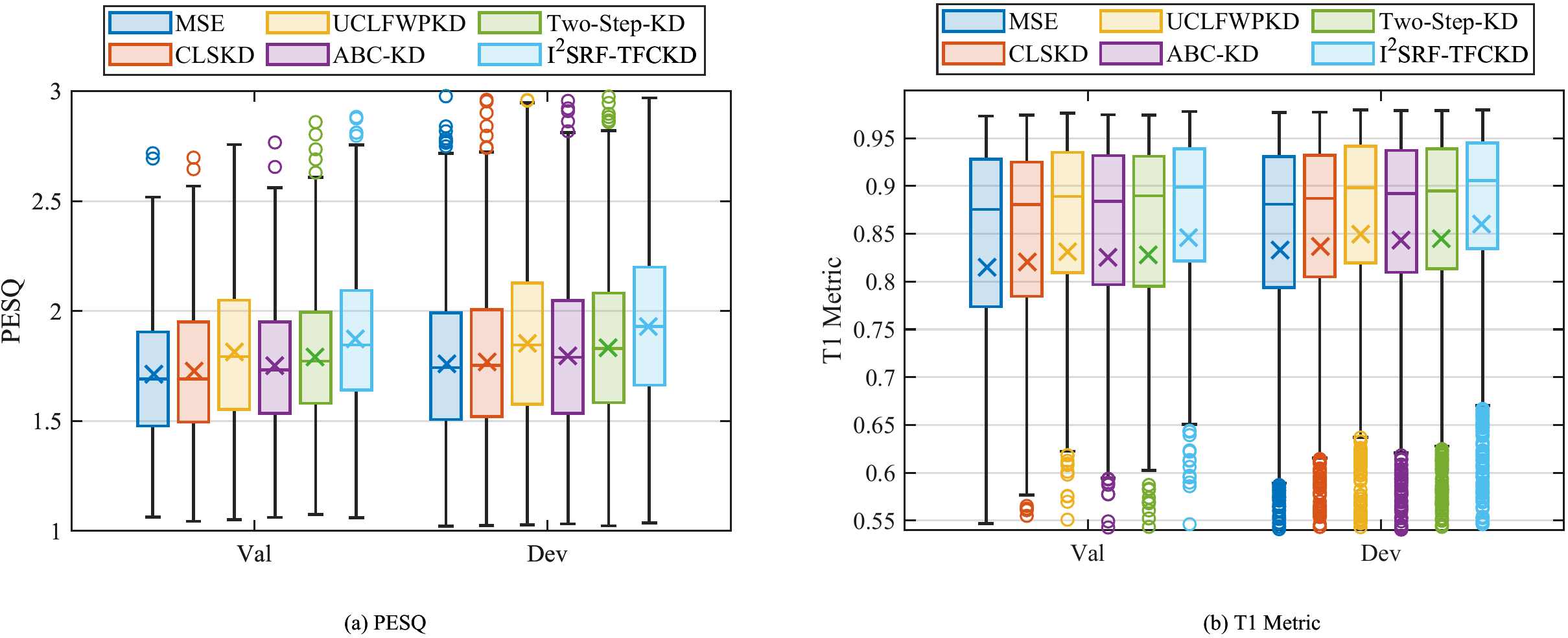}
\caption{Boxplot of PESQ and the T1 Metric for different distillation strategies on the L3DAS23 validation and development sets.}
\label{fig_L3DAS_scatter}
\end{figure}

Fig.\,\ref{fig_L3DAS_scatter} presents a boxplot comparison of representative metric results (PESQ and T1 Metric) for different distillation frameworks on the L3DAS23 validation and development sets. Among all distillation algorithms, the proposed I\textsuperscript{2}SRF-TFCKD method achieves the highest mean values across both metrics. Compared with the second-highest performing UCLFWPKD, the proposed framework has advantages in both the average metric and the concentration degree of high values, further validating the positive contributions of intra-inter set knowledge flow and time-frequency cross-calibration to the overall distillation effect. 

\subsubsection{Metric comparison across models}
\begin{table}[h]
\renewcommand{\arraystretch}{1.1}
\caption{Comparison of objective speech evaluation metrics between distilled and undistilled models on the development set in the SE track of L3DAS23.}
\label{table_result_l3das}
\centering
\begin{threeparttable}
	\resizebox{\linewidth}{!}{
		\begin{tabular}{lcccccccc}
			\toprule
			Methods & Casual \& NFI & Param. (M) & FLOPs (G/s) & RTF & PESQ$\uparrow$ & WER$\downarrow$ & STOI$\uparrow$ & T1 Metric$\uparrow$\\
			\toprule
			Noisy$^1$ & - & - & - & - & 1.166 & 0.474 & 0.575 & 0.550\\ %
			\midrule
			Neural Beamforming~\cite{ren2021neural}$\dagger$ & \XSolidBrush & 5.5 & 32.15 & - & - & 0.569 & 0.684 & 0.557 \\
			EaBNet~\cite{li2022embedding}$\dagger$ & \Checkmark & 2.8 & 7.38 & 0.59 & - & 0.549 & 0.724 & 0.587 \\
			LMFCA-Net~\cite{10889867}$\ddagger$ & \Checkmark & 2.1 & 2.20 & 0.16 & 1.843 & 0.199 & 0.865 & 0.833 \\
			CCA Speech~\cite{wang2023stream}$\dagger$ & \XSolidBrush & - & - & - & - & 0.293 & 0.836 & 0.771 \\
			DeFT-AN~\cite{lee2023deft}$\ddagger$ & \XSolidBrush & 2.7 & 37.99  & - & 1.976 & 0.137 & 0.866 & 0.864 \\
			DPDCRN-T & \Checkmark & 3.5 & 13.71 & 0.36 & 2.036 & 0.129 & 0.885 & 0.878 \\
			DPDCRN-S & \Checkmark & 0.6 & 2.44 & 0.09 & 1.711 & 0.185 & 0.850 & 0.832\\
			\hline
			\hline
			\addlinespace[4pt]
			KD methods &  &  &  &  &  & &\\
			\cline{1-1}
			\addlinespace[4pt]
			-CLSKD~\cite{cheng2022cross}$\dagger$ & \Checkmark & 0.6 & 2.44 & 0.09 & 1.768 & 0.178 & 0.851 & 0.837\\
			-UCLFWPKD~\cite{cheng2024residual}$\dagger$ & \Checkmark & 0.6 & 2.44 & 0.09 & 1.852 & 0.162 & 0.861 & 0.850 \\
			-ABC-KD~\cite{wan23_interspeech}$\ddagger$ & \Checkmark & 0.6 & 2.44 & 0.09 & 1.795 & 0.169 & 0.856 & 0.843 \\
			-Two-Step KD~\cite{nathoo2024two}$\ddagger$ & \Checkmark & 0.6 & 2.44 & 0.09 & 1.832 & 0.169 & 0.859 & 0.845 \\
			-MPMTNet-KD~\cite{11029670}$\ddagger$ & \Checkmark & 0.6 & 2.44 & 0.09 & 1.844 & 0.165 & 0.860 & 0.847 \\
			-M1 & \Checkmark & 0.6 & 2.44 & 0.09 & 1.774 & 0.179 & 0.850 & 0.836 \\
			-M2 & \Checkmark & 0.6 & 2.44 & 0.09 & 1.851 & 0.161 & 0.859 & 0.849 \\
			-M3 & \Checkmark & 0.6 & 2.44 & 0.09 & 1.914 & 0.152 & 0.866 & 0.857 \\
			-M4 (prop.) & \Checkmark & 0.6 & 2.44 & 0.09 & 1.929 & 0.150 & 0.870 & 0.860 \\
			
			\bottomrule
	\end{tabular}}
	\begin{tablenotes}
		\fontsize{6pt}{7pt}\selectfont
		\item[1] Note: Noisy data is taken from the first channel of microphone A.
	\end{tablenotes}
\end{threeparttable}
\end{table}

Table\,\ref{table_result_l3das} presents objective metric results on the L3DAS23 development set, comparing multi-channel SE models with and without KD. The base model DPDCRN achieved first place on the blind test set of the L3DAS23 challenge and is causally adjusted in this paper. Application of various distillation strategies to the student results in varying degrees of improvement across metrics. The proposed I\textsuperscript{2}SRF-TFCKD method achieves optimal distillation effect, particularly in terms of PESQ and WER, benefiting from designs tailored to the intrinsic characteristics of the SE task. In comparisons with other SE frameworks, the proposed distilled DPDCRN-S outperforms the recently proposed multi-channel real-time framework LMFCA-Net across all metrics, whereas the situation is reversed before distillation. This indicates that the proposed distillation strategy fully exploits the latent potential of the student model. Although a performance gap still exists between the student model and the DeFT-AN, the latter is constrained by its non-casual inference paradigm and extremely high FLOPs. Overall, the distilled student model maintains causal inference with low computational overhead while achieving competitive performance compared to SOTA SE methods in enhancement results.

\section{Conclusion}
\label{section_conclusion}

In this paper, we propose an intra-inter set distillation framework with time-frequency calibration (I\textsuperscript{2}SRF-TFCKD) for SE model compression. Specifically, I\textsuperscript{2}SRF-TFCKD partitions intermediate layers of SE models into multiple correlated sets. Within each set, intra-set cross-layer knowledge transfer is performed, and representative features are generated through residual fusion. The representative features of multiple correlated sets are then integrated into a fused feature set to enable inter-set distillation. Furthermore, we design a multi-layer time-frequency cross-calibration distillation tailored to SE tasks: the temporal stream captures importance distributions across speech frames, while the spectral stream focuses on distillation contributions of layer-wise spectral features. Experimental results on two public datasets (DNS and L3DAS23) demonstrate that the proposed distillation framework stably and effectively improves the enhancement performance of the low-complexity student model, narrows the gap with the teacher and outperforms other distillation strategies across various metrics. In future work, we will further investigate multi-teacher distillation for speech models.



\bibliographystyle{IEEEtran}
\bibliography{main}

\begin{thebibliography}{10}
\providecommand{\url}[1]{#1}
\csname url@samestyle\endcsname
\providecommand{\newblock}{\relax}
\providecommand{\bibinfo}[2]{#2}
\providecommand{\BIBentrySTDinterwordspacing}{\spaceskip=0pt\relax}
\providecommand{\BIBentryALTinterwordstretchfactor}{4}
\providecommand{\BIBentryALTinterwordspacing}{\spaceskip=\fontdimen2\font plus
\BIBentryALTinterwordstretchfactor\fontdimen3\font minus
  \fontdimen4\font\relax}
\providecommand{\BIBforeignlanguage}[2]{{%
\expandafter\ifx\csname l@#1\endcsname\relax
\typeout{** WARNING: IEEEtran.bst: No hyphenation pattern has been}%
\typeout{** loaded for the language `#1'. Using the pattern for}%
\typeout{** the default language instead.}%
\else
\language=\csname l@#1\endcsname
\fi
#2}}
\providecommand{\BIBdecl}{\relax}
\BIBdecl

\bibitem{zheng2023sixty}
C.~Zheng, H.~Zhang, W.~Liu, X.~Luo, A.~Li, X.~Li, and B.~C. Moore, ``Sixty
  years of frequency-domain monaural speech enhancement: From traditional to
  deep learning methods,'' \emph{Trends in Hearing}, vol.~27, pp. 1--52, 2023.

\bibitem{reddy20_interspeech}
C.~K. Reddy, V.~Gopal, R.~Cutler, E.~Beyrami, R.~Cheng, H.~Dubey,
  S.~Matusevych, R.~Aichner, A.~Aazami, S.~Braun, P.~Rana, S.~Srinivasan, and
  J.~Gehrke, ``The interspeech 2020 deep noise suppression challenge: Datasets,
  subjective testing framework, and challenge results,'' in \emph{Interspeech
  2020}, 2020, pp. 2492--2496.

\bibitem{ZHU2025107805}
X.~Zhu, J.~Yang, Y.~Yang, W.~Tu, and Z.~Wang, ``Lightweight real-time speech
  enhancement: State-space models and multi-spectral scanning techniques,''
  \emph{Neural Networks}, vol. 191, p. 107805, 2025.

\bibitem{tang2024direct}
J.~Tang, S.~Chen, G.~Niu, H.~Zhu, J.~T. Zhou, C.~Gong, and M.~Sugiyama,
  ``Direct distillation between different domains,'' in \emph{European
  Conference on Computer Vision}.\hskip 1em plus 0.5em minus 0.4em\relax
  Springer, 2024, pp. 154--172.

\bibitem{gong2025beyond}
G.~Gong, J.~Wang, J.~Xu, D.~Xiang, Z.~Zhang, L.~Shen, Y.~Zhang, J.~JunhuaShu,
  Z.~ZhaolongXing, Z.~Chen \emph{et~al.}, ``Beyond logits: Aligning feature
  dynamics for effective knowledge distillation,'' in \emph{Proceedings of the
  63rd Annual Meeting of the Association for Computational Linguistics (Volume
  1: Long Papers)}, 2025, pp. 23\,067--23\,077.

\bibitem{11029670}
W.~Zhou, B.~Jian, Y.~Liu, and Q.~Jiang, ``Multiattentive perception and
  multilayer transfer network using knowledge distillation for rgb-d indoor
  scene parsing,'' \emph{IEEE Transactions on Neural Networks and Learning
  Systems}, vol.~36, no.~10, pp. 18\,287--18\,299, 2025.

\bibitem{wan23_interspeech}
Y.~Wan, Y.~Zhou, X.~Peng, K.-W. Chang, and Y.~Lu, ``Abc-kd:
  Attention-based-compression knowledge distillation for deep learning-based
  noise suppression,'' in \emph{Interspeech 2023}, 2023, pp. 2528--2532.

\bibitem{nathoo2024two}
R.~D. Nathoo, M.~Kegler, and M.~Stamenovic, ``Two-step knowledge distillation
  for tiny speech enhancement,'' in \emph{ICASSP 2024-2024 IEEE International
  Conference on Acoustics, Speech and Signal Processing (ICASSP)}.\hskip 1em
  plus 0.5em minus 0.4em\relax IEEE, 2024, pp. 10\,141--10\,145.

\bibitem{cheng2024residual}
J.~Cheng, R.~Liang, L.~Zhou, L.~Zhao, C.~Huang, and B.~W. Schuller, ``Residual
  fusion probabilistic knowledge distillation for speech enhancement,''
  \emph{IEEE/ACM Transactions on Audio, Speech, and Language Processing},
  vol.~32, pp. 2680--2691, 2024.

\bibitem{cheng2023dual}
J.~Cheng, C.~Pang, R.~Liang, J.~Fan, and L.~Zhao, ``Dual-path dilated
  convolutional recurrent network with group attention for multi-channel speech
  enhancement,'' in \emph{ICASSP 2023-2023 IEEE International Conference on
  Acoustics, Speech and Signal Processing (ICASSP)}.\hskip 1em plus 0.5em minus
  0.4em\relax IEEE, 2023, pp. 1--2.

\bibitem{FAN2023508}
C.~Fan, H.~Zhang, A.~Li, W.~Xiang, C.~Zheng, Z.~Lv, and X.~Wu, ``Compnet:
  Complementary network for single-channel speech enhancement,'' \emph{Neural
  Networks}, vol. 168, pp. 508--517, 2023.

\bibitem{hu20g_interspeech}
Y.~Hu, Y.~Liu, S.~Lv, M.~Xing, S.~Zhang, Y.~Fu, J.~Wu, B.~Zhang, and L.~Xie,
  ``Dccrn: Deep complex convolution recurrent network for phase-aware speech
  enhancement,'' in \emph{Interspeech 2020}, 2020, pp. 2472--2476.

\bibitem{zhao2022frcrn}
S.~Zhao, B.~Ma, K.~N. Watcharasupat, and W.-S. Gan, ``Frcrn: Boosting feature
  representation using frequency recurrence for monaural speech enhancement,''
  in \emph{ICASSP 2022-2022 IEEE international conference on acoustics, speech
  and signal processing (ICASSP)}.\hskip 1em plus 0.5em minus 0.4em\relax IEEE,
  2022, pp. 9281--9285.

\bibitem{le21b_interspeech}
X.~Le, H.~Chen, K.~Chen, and J.~Lu, ``Dpcrn: Dual-path convolution recurrent
  network for single channel speech enhancement,'' in \emph{Interspeech 2021},
  2021, pp. 2811--2815.

\bibitem{NI2026103726}
Y.~Ni, R.~Liang, X.~Hao, J.~Cheng, Q.~Wang, C.~Huang, C.~Zou, W.~Zhou, W.~Ding,
  and B.~W. Schuller, ``Affine modulation-based audiogram fusion network for
  joint noise reduction and hearing loss compensation,'' \emph{Information
  Fusion}, vol. 127, p. 103726, 2026.

\bibitem{xia2020weighted}
Y.~Xia, S.~Braun, C.~K. Reddy, H.~Dubey, R.~Cutler, and I.~Tashev, ``Weighted
  speech distortion losses for neural-network-based real-time speech
  enhancement,'' in \emph{ICASSP 2020-2020 IEEE International Conference on
  Acoustics, Speech and Signal Processing (ICASSP)}.\hskip 1em plus 0.5em minus
  0.4em\relax IEEE, 2020, pp. 871--875.

\bibitem{9414965}
T.~Vuong, Y.~Xia, and R.~M. Stern, ``A modulation-domain loss for
  neural-network-based real-time speech enhancement,'' in \emph{ICASSP 2021 -
  2021 IEEE International Conference on Acoustics, Speech and Signal Processing
  (ICASSP)}, 2021, pp. 6643--6647.

\bibitem{hinton2015distilling}
G.~Hinton, O.~Vinyals, and J.~Dean, ``Distilling the knowledge in a neural
  network,'' \emph{arXiv preprint arXiv:1503.02531}, 2015.

\bibitem{yoon2023inter}
J.~W. Yoon, B.~J. Woo, S.~Ahn, H.~Lee, and N.~S. Kim, ``Inter-kd: Intermediate
  knowledge distillation for ctc-based automatic speech recognition,'' in
  \emph{2022 IEEE Spoken Language Technology Workshop (SLT)}.\hskip 1em plus
  0.5em minus 0.4em\relax IEEE, 2023, pp. 280--286.

\bibitem{9767637}
R.~Liu, B.~Sisman, G.~Gao, and H.~Li, ``Decoding knowledge transfer for neural
  text-to-speech training,'' \emph{IEEE/ACM Transactions on Audio, Speech, and
  Language Processing}, vol.~30, pp. 1789--1802, 2022.

\bibitem{hao20b_interspeech}
X.~Hao, S.~Wen, X.~Su, Y.~Liu, G.~Gao, and X.~Li, ``Sub-band knowledge
  distillation framework for speech enhancement,'' in \emph{Interspeech 2020},
  2020, pp. 2687--2691.

\bibitem{cheng2022cross}
J.~Cheng, R.~Liang, Y.~Xie, L.~Zhao, B.~Schuller, J.~Jia, and Y.~Peng,
  ``Cross-layer similarity knowledge distillation for speech enhancement.'' in
  \emph{INTERSPEECH}, 2022, pp. 926--930.

\bibitem{CHEN2025103218}
H.~Chen, C.~Wang, Q.~Wang, J.~Du, S.~M. Siniscalchi, G.~Wan, J.~Pan, and
  H.~Ding, ``Cross-attention among spectrum, waveform and ssl representations
  with bidirectional knowledge distillation for speech enhancement,''
  \emph{Information Fusion}, vol. 122, p. 103218, 2025.

\bibitem{10890458}
X.~Yuan, S.~Liu, H.~Chen, L.~Zhou, J.~Li, and J.~Hu, ``Dynamic
  frequency-adaptive knowledge distillation for speech enhancement,'' in
  \emph{ICASSP 2025 - 2025 IEEE International Conference on Acoustics, Speech
  and Signal Processing (ICASSP)}, 2025, pp. 1--5.

\bibitem{han2024distil}
R.~Han, W.~Xu, Z.~Zhang, M.~Liu, and L.~Xie, ``Distil-dccrn: A small-footprint
  dccrn leveraging feature-based knowledge distillation in speech
  enhancement,'' \emph{IEEE Signal Processing Letters}, 2024.

\bibitem{10599808}
A.~Wu, J.~Yu, Y.~Wang, and C.~Deng, ``Prototype-decomposed knowledge
  distillation for learning generalized federated representation,'' \emph{IEEE
  Transactions on Multimedia}, vol.~26, pp. 10\,991--11\,002, 2024.

\bibitem{wang2022semckd}
C.~Wang, D.~Chen, J.-P. Mei, Y.~Zhang, Y.~Feng, and C.~Chen, ``Semckd: Semantic
  calibration for cross-layer knowledge distillation,'' \emph{IEEE Transactions
  on Knowledge and Data Engineering}, vol.~35, no.~6, pp. 6305--6319, 2022.

\bibitem{chen2022fullsubnet+}
J.~Chen, Z.~Wang, D.~Tuo, Z.~Wu, S.~Kang, and H.~Meng, ``Fullsubnet+: Channel
  attention fullsubnet with complex spectrograms for speech enhancement,'' in
  \emph{ICASSP 2022-2022 IEEE international conference on acoustics, speech and
  signal processing (ICASSP)}.\hskip 1em plus 0.5em minus 0.4em\relax IEEE,
  2022, pp. 7857--7861.

\bibitem{defossez20_interspeech}
A.~Défossez, G.~Synnaeve, and Y.~Adi, ``Real time speech enhancement in the
  waveform domain,'' in \emph{Interspeech 2020}, 2020, pp. 3291--3295.

\bibitem{westhausen20_interspeech}
N.~L. Westhausen and B.~T. Meyer, ``Dual-signal transformation lstm network for
  real-time noise suppression,'' in \emph{Interspeech 2020}, 2020, pp.
  2477--2481.

\bibitem{10448310}
X.~Rong, T.~Sun, X.~Zhang, Y.~Hu, C.~Zhu, and J.~Lu, ``Gtcrn: A speech
  enhancement model requiring ultralow computational resources,'' in
  \emph{ICASSP 2024 - 2024 IEEE International Conference on Acoustics, Speech
  and Signal Processing (ICASSP)}, 2024, pp. 971--975.

\bibitem{11464802}
S.~Ahn, J.~Han, B.~J. Woo, and N.~Soo~Kim, ``Fastenhancer: Speed-optimized
  streaming neural speech enhancement,'' in \emph{ICASSP 2026 - 2026 IEEE
  International Conference on Acoustics, Speech and Signal Processing
  (ICASSP)}, 2026, pp. 16\,847--16\,851.

\bibitem{hao2021fullsubnet}
X.~Hao, X.~Su, R.~Horaud, and X.~Li, ``Fullsubnet: A full-band and sub-band
  fusion model for real-time single-channel speech enhancement,'' in
  \emph{ICASSP 2021-2021 IEEE International Conference on Acoustics, Speech and
  Signal Processing (ICASSP)}.\hskip 1em plus 0.5em minus 0.4em\relax IEEE,
  2021, pp. 6633--6637.

\bibitem{li2021two}
A.~Li, W.~Liu, C.~Zheng, C.~Fan, and X.~Li, ``Two heads are better than one: A
  two-stage complex spectral mapping approach for monaural speech
  enhancement,'' \emph{IEEE/ACM Transactions on Audio, Speech, and Language
  Processing}, vol.~29, pp. 1829--1843, 2021.

\bibitem{ren2021neural}
X.~Ren, L.~Chen, X.~Zheng, C.~Xu, X.~Zhang, C.~Zhang, L.~Guo, and B.~Yu, ``A
  neural beamforming network for b-format 3d speech enhancement and
  recognition,'' in \emph{2021 IEEE 31st International Workshop on Machine
  Learning for Signal Processing (MLSP)}.\hskip 1em plus 0.5em minus
  0.4em\relax IEEE, 2021, pp. 1--6.

\bibitem{li2022embedding}
A.~Li, W.~Liu, C.~Zheng, and X.~Li, ``Embedding and beamforming: All-neural
  causal beamformer for multichannel speech enhancement,'' in \emph{ICASSP
  2022-2022 IEEE international conference on acoustics, speech and signal
  processing (ICASSP)}.\hskip 1em plus 0.5em minus 0.4em\relax IEEE, 2022, pp.
  6487--6491.

\bibitem{10889867}
Y.~Zhang, H.~Pei, W.~Wang, and G.~Huang, ``Lmfca-net: A lightweight model for
  multi-channel speech enhancement with efficient narrow-band and cross-band
  attention,'' in \emph{ICASSP 2025 - 2025 IEEE International Conference on
  Acoustics, Speech and Signal Processing (ICASSP)}, 2025, pp. 1--5.

\bibitem{wang2023stream}
H.~Wang, Y.~Fu, J.~Li, M.~Ge, L.~Wang, and X.~Qian, ``Stream attention based
  u-net for l3das23 challenge,'' in \emph{ICASSP 2023-2023 IEEE International
  Conference on Acoustics, Speech and Signal Processing (ICASSP)}.\hskip 1em
  plus 0.5em minus 0.4em\relax IEEE, 2023, pp. 1--2.

\bibitem{lee2023deft}
D.~Lee and J.-W. Choi, ``Deft-an: Dense frequency-time attentive network for
  multichannel speech enhancement,'' \emph{IEEE Signal Processing Letters},
  vol.~30, pp. 155--159, 2023.

\bibitem{fisher1970statistical}
R.~A. Fisher, ``Statistical methods for research workers,'' in
  \emph{Breakthroughs in statistics: Methodology and distribution}.\hskip 1em
  plus 0.5em minus 0.4em\relax Springer, 1970, pp. 66--70.

\end{thebibliography}
%

\begin{IEEEbiography}[{\includegraphics[width=1in,height=1.25in,clip,keepaspectratio]{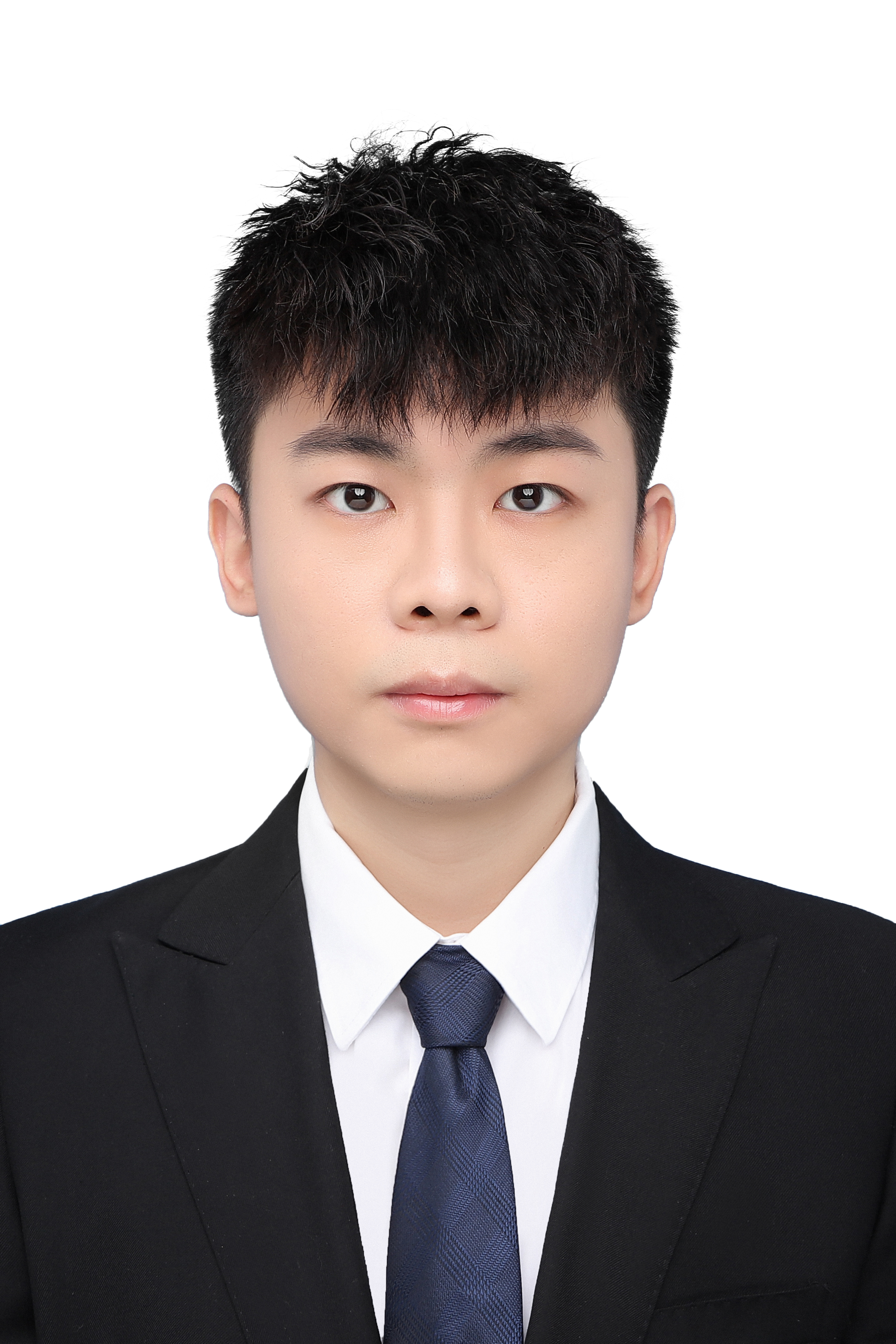}}]{Jiaming Cheng}
	received the PhD degree from Southeast University, Nanjing, China, in 2024. He is currently a Lecturer with the School of Computer Science, Nanjing Audit University, Nanjing, China. His research interests include single/multi-microphone speech processing, hearing aids, transfer learning, and knowledge distillation.
\end{IEEEbiography}
\vspace{-10.0mm}

\begin{IEEEbiography}[{\includegraphics[width=1in,height=1.25in,clip,keepaspectratio]{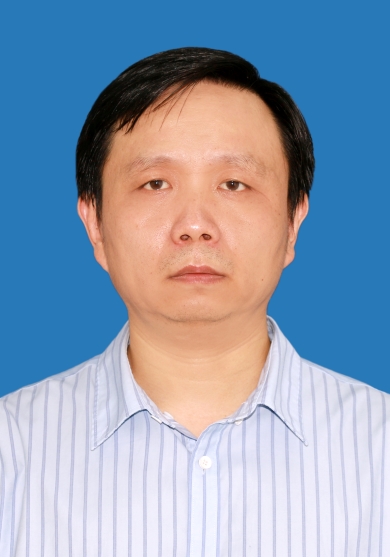}}]{Ruiyu Liang}
	(Member, IEEE) received the PhD degree from Southeast University, China, in 2012. He is currently a Professor with Nanjing Institute of Technology, Nanjing, Jiangsu province, China. His research interests include speech signal processing and signal processing for hearing aids.
\end{IEEEbiography}
\vspace{-10.0mm}

\begin{IEEEbiography}[{\includegraphics[width=1in,height=1.25in,clip,keepaspectratio]{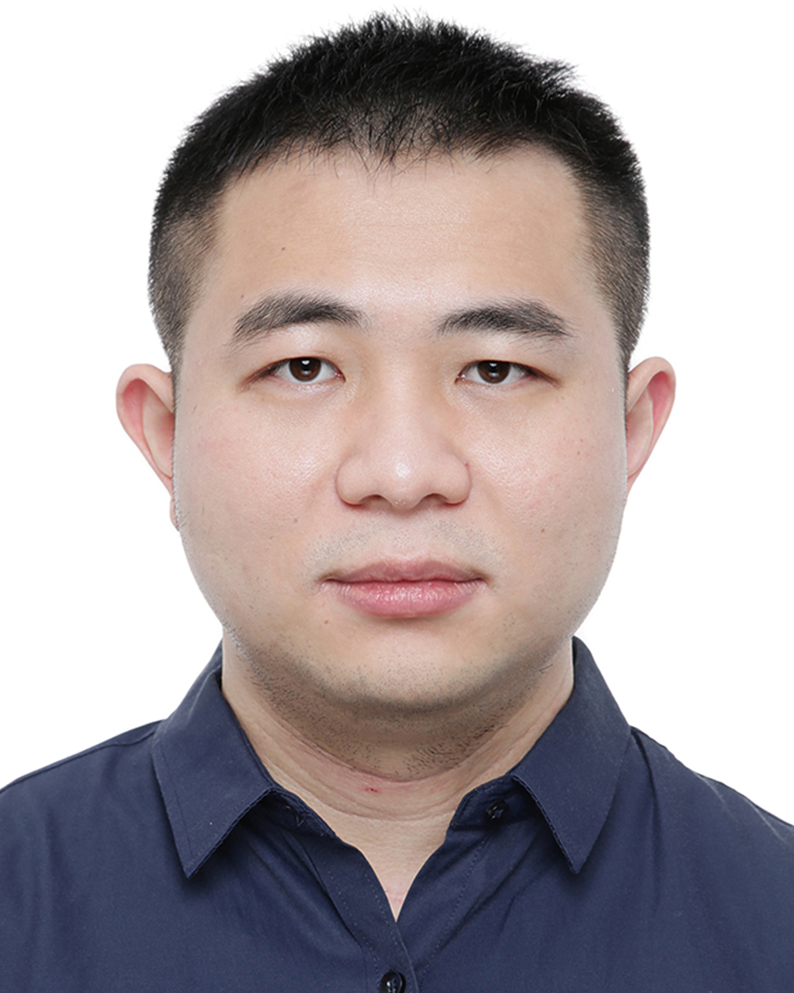}}]{Ye Ni}
	received the M.S.\ degree from Nanjing University, Nanjing, China, in 2022. He is currently working toward a PhD degree from Southeast University, Nanjing, China. His research interests include deep learning-based speech enhancement, acoustic echo cancellation, and signal processing.
\end{IEEEbiography}
\vspace{-10.0mm}

\begin{IEEEbiography}[{\includegraphics[width=1in,height=1.25in,clip,keepaspectratio]{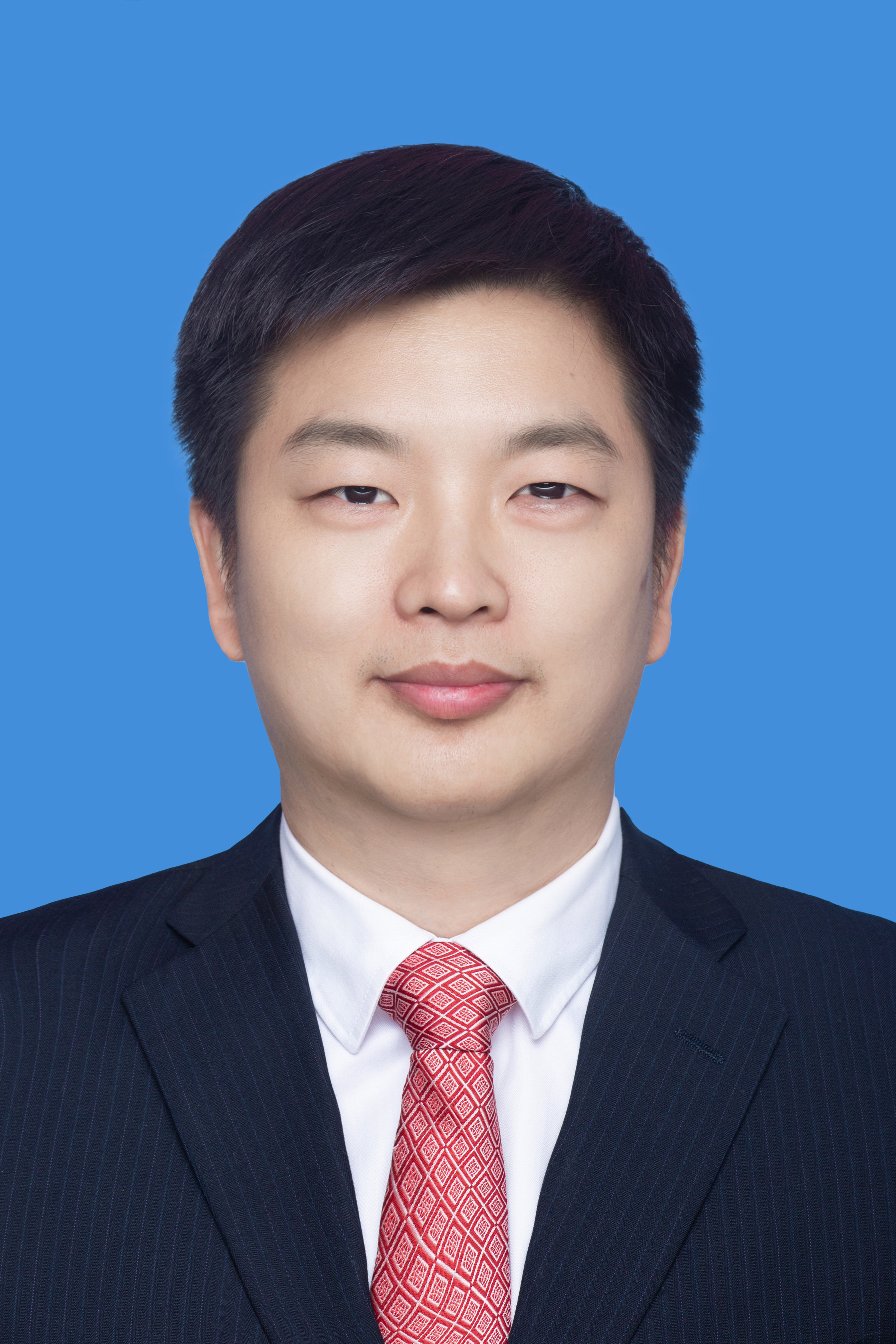}}]{Chao Xu}
	received the PhD degree from the School of Computer, Wuhan University, Wuhan, China, in 2014. He is currently a Professor with the School of Computer Science, Nanjing Audit University, Nanjing, China. His research interests include big data technology and artificial intelligence.
\end{IEEEbiography}
\vspace{-10.0mm}

\begin{IEEEbiography}[{\includegraphics[width=1in,height=1.25in,clip,keepaspectratio]{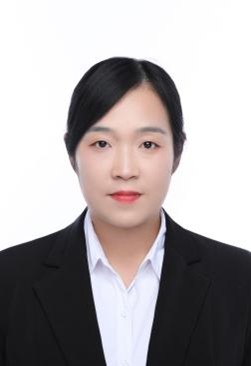}}]{JingLi}
	was born in Xuzhou, Jiangsu Province, China. She obtained her Ph.D. degree in Information and Communication Engineering from Southeast University. Currently, she holds the position of Associate Professor at Nanjing Audit University and concurrently serves as a Researcher at the Post-Doctoral Mobile Research Station in Power Engineering of Southeast University. Her primary research interests focus on signal processing, artificial intelligence, and modeling, with specific applications in acoustic emission signal recognition. 
\end{IEEEbiography}
\vspace{-10.0mm}

\begin{IEEEbiography}[{\includegraphics[width=1in,height=1.25in,clip,keepaspectratio]{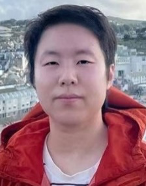}}]{Wei Zhou}
	(IEEE Senior Member) is an Assistant Professor at Cardiff University, United Kingdom. Previously, Wei studied and worked at other institutions such as the University of Waterloo (Canada), the National Institute of Informatics (Japan), the University of Science and Technology of China, Intel, Microsoft Research, and Alibaba Group. Dr Zhou is now an Associate Editor of IEEE Transactions on Neural Networks and Learning Systems (TNNLS), ACM Transactions on Multimedia Computing, Communications, and Applications (TOMM), and Pattern Recognition. Wei’s research interests span multimedia computing, perceptual image processing, and computational vision.
\end{IEEEbiography}
\vspace{-10.0mm}

\begin{IEEEbiography}[{\includegraphics[width=1in,height=1.25in,clip,keepaspectratio]{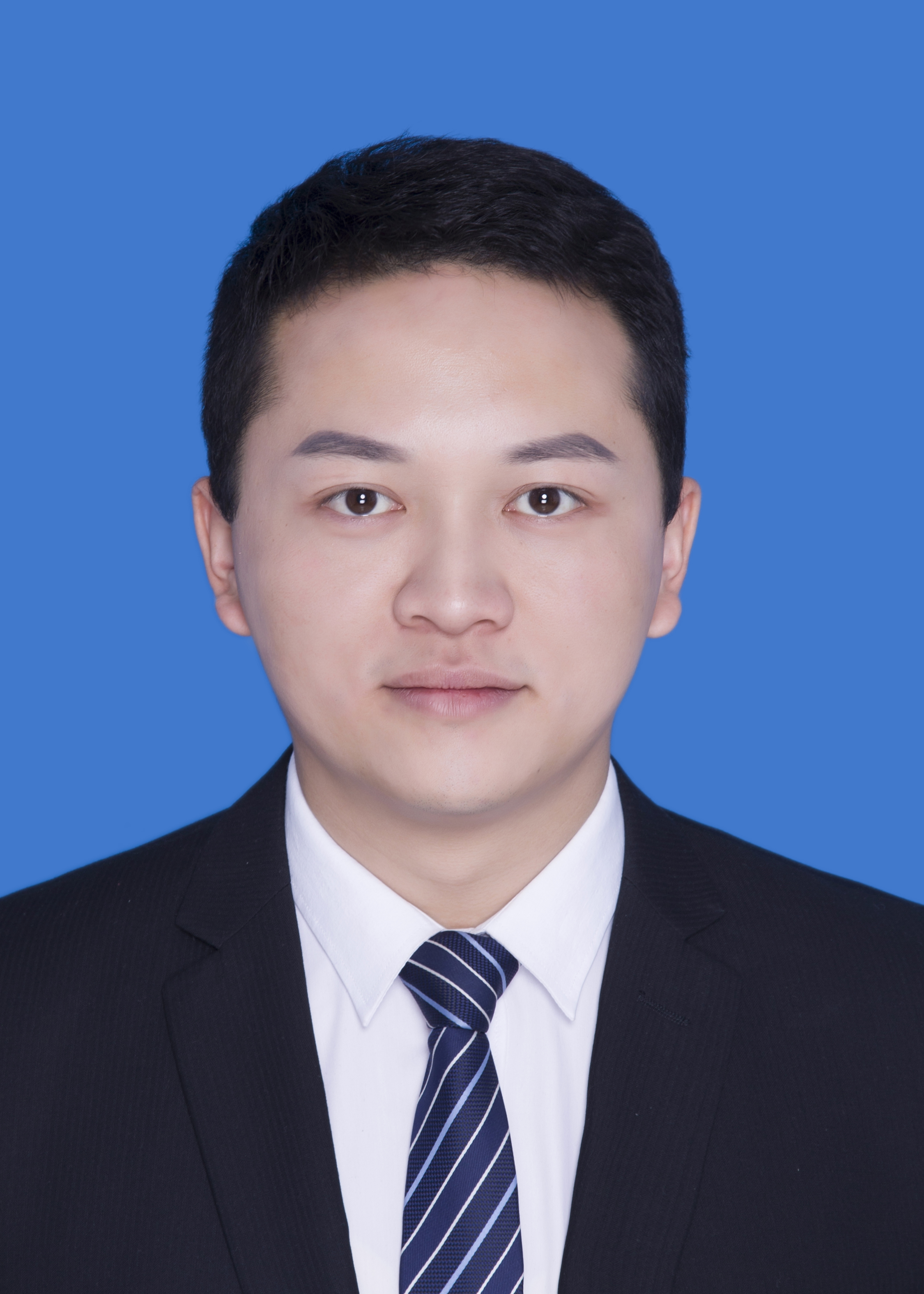}}]{Rui Liu}
	(IEEE Member) is currently a professor in National and Local Joint Engineering Research Center of Mongolian Intelligent Information Processing, Inner Mongolia University. He was the recipient of the “Best Paper Award” at the 2021\&2025 International Conference on Asian Language Processing (IALP). His publications include top-tier NLP/ML/AI conferences and journals, including IEEE-TASLPRO, IEEE-TAFFC, Neural Networks, AAAI, ACMMM, ACL, EMNLP, ICASSP, COLING, INTERSPEECH, etc. He is a member of ISCA, CAAI, CIPS and CCF, and serves as the reviewer for many major referred journal and conference papers. His research interests broadly lie in multilingual human-machine speech interaction.
\end{IEEEbiography}
\vspace{-10.0mm}

\begin{IEEEbiography}[{\includegraphics[width=1in,height=1.25in,clip,keepaspectratio]{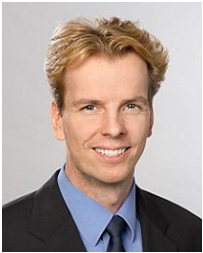}}]{Bj\"orn Schuller}
	(IEEE Fellow) received the diploma, the doctoral degree, and the habilitation and Adjunct Teaching Professorship in the subject area of signal processing and machine intelligence, all in electrical engineering and information technology from the Technical University of Munich (TUM), Germany, in 1999, 2006, and 2012, respectively. He is a Full Professor of artificial intelligence, and the Head of GLAM -- the Group on Language, Audio \& Music, Imperial College London, U.K., a Full Professor and Chair of Health Informatics at TUM, co-founding CEO and current CSO of audEERING. He is also with Munich's MCML, MDSI, and MIBE. Previous stations include the University of Augsburg and Passau, Germany, as Full Professor, the French CNRS, and Joanneum Research in Graz, Austria. He is  President Emeritus and Fellow of the AAAC, Fellow of the ACM, BCS, DIRDI, ELLIS, IEEE, and ISCA. He authored or coauthored five books and more than 1500 publications in peer reviewed books, journals, and conference proceedings leading to more than 65k citations (h-index = 116).
\end{IEEEbiography}
\vspace{-10.0mm}

\begin{IEEEbiography}[{\includegraphics[width=1in,height=1.25in,clip,keepaspectratio]{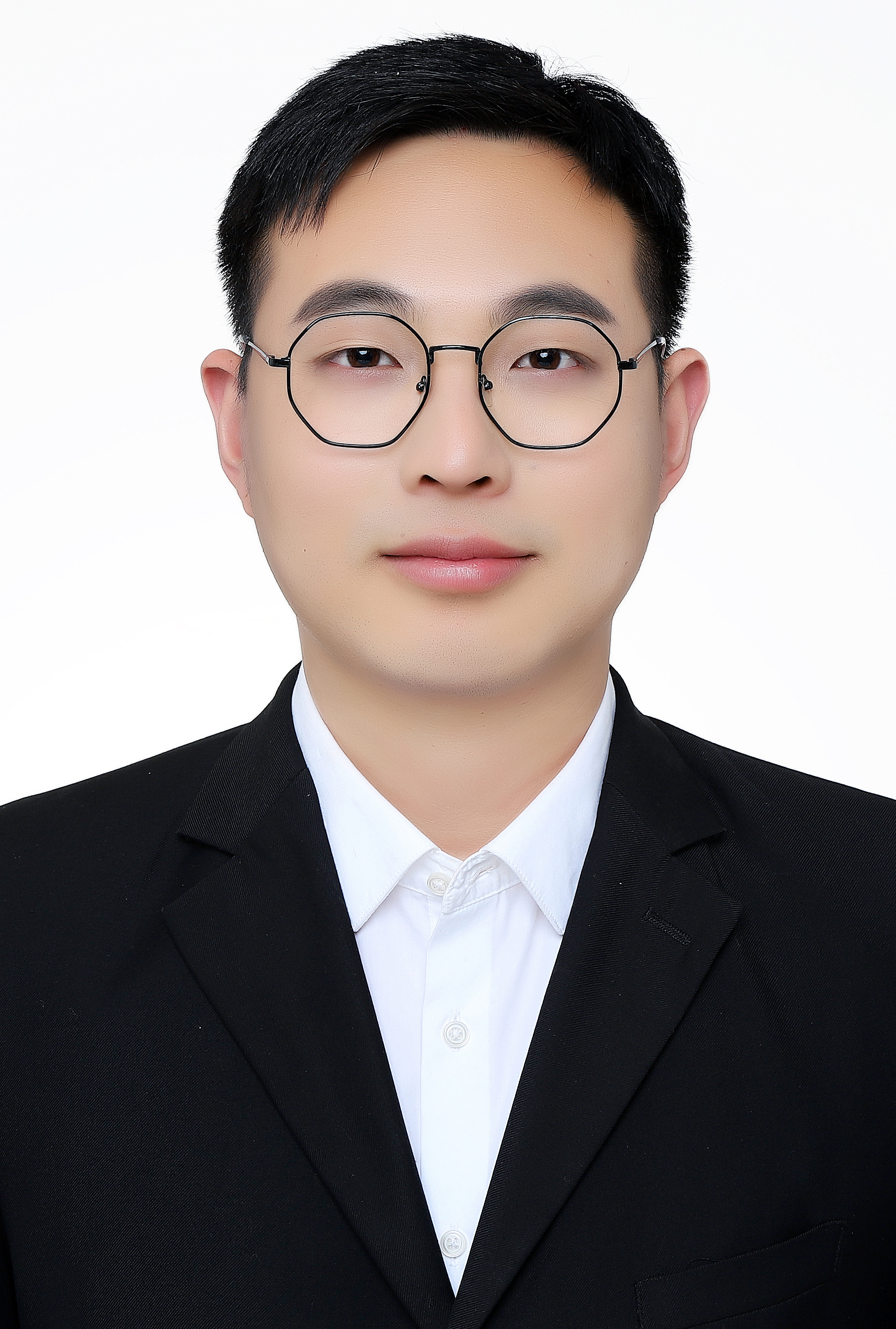}}]{Xiaoshuai Hao}
	received his Ph.D. from the Institute of Information Engineering, Chinese Academy of Sciences, in 2023. He is currently a research expert in multimodal algorithms for autonomous driving and robotics. His research interests include embodied intelligence, multimodal learning, and autonomous driving. Dr. Hao has published over 50 papers in top-tier journals and conferences, such as TIP, Information Fusion, NeurIPS, ICLR, ICML, CVPR, ICCV, ECCV, ACL, AAAI, and ICRA. He has achieved remarkable success in international competitions, securing top-three placements at prestigious conferences like CVPR and ICCV. Additionally, he serves on the editorial board of Data Intelligence and is an organizer for the RoDGE Workshop at ICCV 2025 and the RoboSense Challenge at IROS 2025.
\end{IEEEbiography}

\end{document}